\documentclass[superscriptaddress,amsmath,amssymb,aps,pre]{revtex4-2}
\usepackage[T1]{fontenc}
\usepackage{lmodern}
\usepackage[english]{babel}
\usepackage{amsmath,amssymb,bm,mathtools}
\usepackage{graphicx}
\usepackage{booktabs}
\usepackage{xcolor}
\usepackage{hyperref}
\usepackage{microtype}

\hypersetup{
  colorlinks=true,
  linkcolor=blue,
  citecolor=blue,
  urlcolor=blue
}

\begin{document}
\title{Diffusion with stochastic resetting in the presence of a delta killing trap: Drift-controlled survival regimes}
\author{A. Mazzolo}
\affiliation{Universit\'e Paris-Saclay, CEA, Service d'\'Etudes des R\'eacteurs et de Math\'ematiques Appliqu\'ees, 91191, Gif-sur-Yvette, France}
\email{alain.mazzolo@cea.fr}
\begin{abstract}
We study a one-dimensional diffusive particle subject to stochastic resetting to its initial position, in the presence of an imperfect, localized target that can absorb (kill) the particle, modeled by a delta-function killing rate. The central question is how stochastic resetting competes with drift-induced transience and the target's finite reactivity. Exact Laplace-space expressions are derived for the non-normalized propagator and survival probability, for arbitrary diffusion coefficient $D$, resetting rate $r$, and killing strength $k$, first in the absence of drift and then under a constant drift. The long-time behavior separates into distinct regimes. Without resetting, unbiased diffusion exhibits the recurrent algebraic survival law $S(t)\sim t^{-1/2}$, whereas any nonzero drift renders the motion transient with respect to the target and leaves a nonzero ultimate survival probability. By contrast, any $r>0$ 
and $k>0$ repeatedly renews encounters with the target and restores eventual absorption, yielding an exponential survival law $S(t)\sim A e^{-\theta t}$. The decay rate $\theta$ is given by the dominant (rightmost) pole of the Laplace transform, and the surviving density converges after normalization to an explicit quasi-stationary profile. The driftless and perfectly absorbing limits recover standard resetting results. These formulas provide a unified description of the crossover between recurrence-controlled, transience-controlled, and renewal-controlled survival.
\end{abstract}

\maketitle

\section{Introduction}
Stochastic resetting interrupts a random trajectory and restarts it from a prescribed position. Even this simple nonequilibrium mechanism can qualitatively reshape first-passage properties, for example by replacing broad first-passage statistics with a renewal process characterized by a finite time scale \cite{EvansMajumdar2011,EvansMajumdarSchehr2020}. For a generic stochastic process, the same mechanism---usually referred to as restart---was shown to produce a universal coefficient of variation of the completion time at the optimal restart rate \cite{Reuveni2016}, and was subsequently incorporated into a general framework for first-passage processes under restart \cite{PalReuveni2017}. Since the original formulation, resetting protocols have been generalized in many directions. Examples include partial resetting, in which only a fraction of the displacement from the reset point is removed \cite{TalFriedman2022}; resetting by a random amplitude, allowing partial returns or overshoots \cite{Dahlenburg2021}; noninstantaneous stochastic returns generated by switching on an external confining trap \cite{GuptaPlata2021}; and resetting of drift--diffusion processes, where the interplay between advective and diffusive transport, quantified by the P\'eclet number, determines whether resetting accelerates or hinders first passage \cite{RayMondalReuveni2019}. A broad account of these and many other developments appears in the Journal of Physics A focus issue \emph{Stochastic Resetting: Theory and Applications}, published to mark the tenth anniversary of \emph{Diffusion with Stochastic Resetting} \cite{KunduReuveni2024}; see also Ref.~\cite{GuptaJayannavar2022} for a concise pedagogical introduction.

Most of the standard resetting literature treats the target as perfectly absorbing. In many reaction and search problems, however, reaching the target does not imply immediate removal: reaction occurs only after sufficient contact. In one dimension, a localized imperfect target is naturally represented by a delta-function killing term \cite{Taitelbaum}. The driftless version of diffusion with Poissonian resetting and partial absorption at the origin was studied in detail by Whitehouse, Evans, and Majumdar \cite{Whitehouse2013}. They derived the Laplace-space survival probability and its exponential long-time decay, calculated and optimized the mean absorption time, and extended the analysis to the survival of a target in the presence of many independent searchers. The driftless calculations presented below therefore serve both as a reference point and as a reformulation of that earlier problem within a Green-function and renewal framework. In addition to recovering the corresponding survival and mean-absorption-time results, we derive the full non-normalized propagator and the associated quasi-stationary spatial profile.

The principal extension developed here is the inclusion of a constant drift. Without resetting, the fate of the particle is then governed by the competition between recurrence, transience, and finite reactivity. Unbiased one-dimensional diffusion is recurrent, so the local time accumulated at the target diverges and absorption eventually becomes certain despite the finite killing strength. By contrast, any nonzero constant drift makes the motion transient with respect to the fixed target. When the drift points away from the origin, the target may never be reached; when it initially points toward the origin, the particle reaches the target with probability one but eventually drifts past it and accumulates only a finite local time there. In either case, a partially absorbing trap leaves a strictly positive ultimate survival probability. This leads to the central question addressed in this work: \emph{can repeated resetting turn an imperfect local reaction into certain eventual absorption even when drift makes the reset-free motion transient?} We show that it can. Resetting repeatedly returns the particle to the same starting point and renews its opportunities for reactive contact. For every positive resetting rate and killing strength, the drift-induced survival plateau is replaced by an exponential survival law governed by an isolated pole of the renewal denominator and by an associated quasi-stationary profile.

To make the distinction from the earlier driftless study of Whitehouse, Evans, and Majumdar~\cite{Whitehouse2013} explicit, the new elements of the present work are the full non-normalized propagator and associated quasi-stationary profile for the finite-reactivity resetting problem, the extension to a constant drift, the exact characterization of the drift-induced reset-free survival plateau, the demonstration that resetting removes this plateau and restores eventual absorption with exponential decay, and the resulting asymmetric quasi-stationary state in the presence of drift.

We obtain these results from exact Green functions and renewal identities. The article is organized as follows. Section~II revisits the driftless problem, explicitly connecting the present formulation with Ref.~\cite{Whitehouse2013}, and derives the combined resetting--killing propagator, survival probability, long-time decay rate, and quasi-stationary profile. Section~III extends the construction to a constant drift and analyzes both the reset-free survival plateau and the exponential decay restored by resetting. Section~IV summarizes the three resulting survival regimes and their physical interpretation. Appendix~\ref{app:optimal-resetting-rate} returns to the optimization problem first considered in Ref.~\cite{Whitehouse2013}: it introduces a dimensionless reactivity parameter that remains fixed as the resetting rate is varied and uses it to establish directly the existence and uniqueness of the optimal rate. Throughout, the diffusion coefficient $D$ is kept explicit.

\section{Brownian Motion with Stochastic Resetting and a Delta Killing Point}
\begin{figure}[ht]
\centering
\includegraphics[width=4in,height=4in]{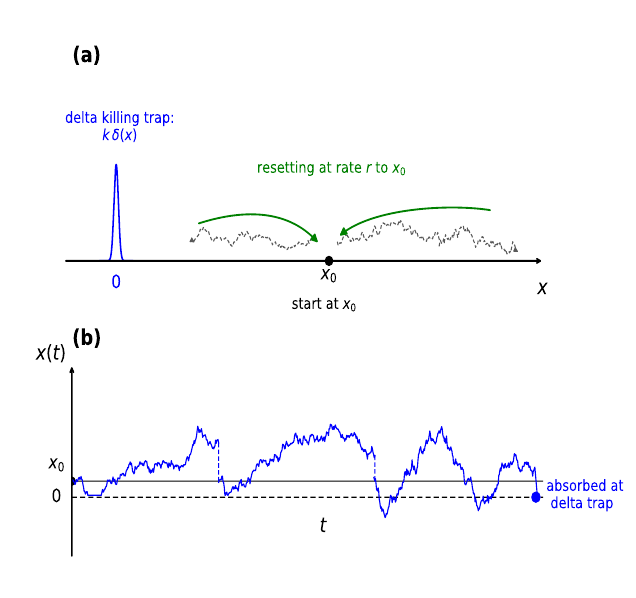}
\setlength{\abovecaptionskip}{15pt}
\caption{Illustration of Brownian motion with stochastic resetting in the
presence of a localized delta killing trap.
(a) A Brownian particle starts at $x_0$ and diffuses freely on the real
line. At random times, drawn from a Poisson process of rate $r$, the
particle is instantaneously reset to its starting position $x_0$. A
fixed, partially absorbing trap of strength $k$, modeled by the delta
killing potential $k \delta(x)$, is located at the origin.
(b) A representative stochastic trajectory as a function of time. Vertical
dashed lines indicate resetting events, and the trajectory ends when the
particle is finally absorbed at the origin. Because the trap is only
partially absorbing, the particle may cross $x=0$ several times before
this final absorption event, reflecting the partial (non-perfect) nature of the localized killing.
}
\label{fig:resetting_delta}
\end{figure}

We consider a one-dimensional Brownian particle with diffusion coefficient $D>0$, reset to its initial position $x_0\in\mathbb R$ at Poisson rate $r\geq0$, and subject to a localized killing term $k\delta(x)$ of strength $k\geq0$ at the origin. With the present convention, $k$ has dimensions of length per unit time: since $\delta(x)$ has dimension $L^{-1}$, the Feynman--Kac exponent $k\int_0^t\delta(X_u)\,\mathrm{d}u$ is dimensionless. These parameters retain the same meaning throughout the article. We first review the two elementary limiting problems---resetting without killing and killing without resetting---and then combine them through a renewal construction. The resulting formulas make transparent the transition from algebraic survival at $r=0$ to exponential survival decay for $r>0$, as well as the perfectly absorbing limit $k\to\infty$. Figure~\ref{fig:resetting_delta} provides a schematic overview of the process.

%%%%%%%%%%%%%%%%%%%%%%%%%%%%%%%%%%%%%%%%%%%%%%%%%%%%%%%%
%%                      FIGURE                        %%
%%%%%%%%%%%%%%%%%%%%%%%%%%%%%%%%%%%%%%%%%%%%%%%%%%%%%%%%

We use the diffusion convention
\begin{equation}
    \mathrm{d} X_t = \sqrt{2D}\,\mathrm{d} W_t,
\end{equation}
where $W_t$ is a Wiener process; with this convention, the free Fokker--Planck generator is $D\partial_x^2$. The free propagator is
\begin{equation}
    G_0(x,t|x_0)
    = \frac{1}{\sqrt{4\pi Dt}}
      \exp\left[-\frac{(x-x_0)^2}{4Dt}\right],
\end{equation}
and its Laplace transform, with Laplace variable $s$, is
\begin{equation}
    \widetilde G_0(x,s|x_0)
    = \int_0^\infty \mathrm{e}^{-st}G_0(x,t|x_0)\,\mathrm{d} t
    = \frac{1}{2\sqrt{Ds}}\,
      \exp\left[-|x-x_0|\sqrt{\frac{s}{D}}\right].
    \label{eq:G0Laplace}
\end{equation}
Equivalently, if
\begin{equation}
    \lambda(s)=\sqrt{\frac{s}{D}},
\end{equation}
then
\begin{equation}
    \widetilde G_0(x,s|x_0)
    = \frac{1}{2D\lambda(s)}\,
      \mathrm{e}^{-\lambda(s)|x-x_0|}.
\end{equation}
The formulas below are written for arbitrary $x_0\in\mathbb R$; when the limit $k\to\infty$ is interpreted as a perfectly absorbing target at the origin, one takes $x_0>0$.

\subsection{Brownian motion with Poissonian resetting, without killing}

Let $p_r(x,t|x_0)$ denote the normalized propagator of a Brownian particle which resets to $x_0$ at rate $r$. It satisfies \cite{EvansMajumdar2011}
\begin{equation}
    \partial_t p_r(x,t|x_0)
    =D\partial_x^2p_r(x,t|x_0)-rp_r(x,t|x_0)+r\delta(x-x_0),
    \label{eq:resetMaster}
\end{equation}
with $p_r(x,0|x_0)=\delta(x-x_0)$. The term $-rp_r$ removes probability density from all positions due to resets, while $r\delta(x-x_0)$ reinjects the same total probability at the reset point.

A renewal decomposition over the time $\tau$ elapsed since the last reset gives \cite{EvansMajumdarSchehr2020}
\begin{equation}
    p_r(x,t|x_0)
    =\mathrm{e}^{-rt}G_0(x,t|x_0)
    +r\int_0^t \mathrm{e}^{-r\tau}G_0(x,\tau|x_0)\,\mathrm{d}\tau .
    \label{eq:resetRenewal}
\end{equation}
The first term corresponds to trajectories with no reset before time $t$; the second term corresponds to trajectories whose last reset occurred at time $t-\tau$.

Taking the Laplace transform of Eq.~\eqref{eq:resetRenewal} gives
\begin{equation}
    \widetilde p_r(x,s|x_0)
    = \left(1+\frac{r}{s}\right)\widetilde G_0(x,s+r|x_0).
\end{equation}
The resulting exact Laplace-space propagator is
\begin{equation}
    \widetilde p_r(x,s|x_0)
    =\frac{s+r}{s}\,
      \frac{1}{2\sqrt{D(s+r)}}
      \exp\left[-|x-x_0|\sqrt{\frac{s+r}{D}}\right] .
    \label{eq:resetLaplace}
\end{equation}

At long times the no-reset contribution in Eq.~\eqref{eq:resetRenewal} dies out and the process reaches the nonequilibrium stationary state
\begin{equation}
    p_r^\mathrm{st}(x|x_0)
    =\frac{\alpha_0}{2}\,\mathrm{e}^{-\alpha_0|x-x_0|},
    \qquad
    \alpha_0=\sqrt{\frac{r}{D}} .
    \label{eq:stationaryReset}
\end{equation}
Equations~\eqref{eq:resetMaster}--\eqref{eq:stationaryReset} are the standard results for diffusion with stochastic resetting. In the notation of Evans and Majumdar~\cite{EvansMajumdar2011}, and of the review by Evans, Majumdar, and Schehr~\cite{EvansMajumdarSchehr2020}, the reset point is often denoted $X_r$; setting $X_r=x_0$ gives exactly the same master equation, renewal representation, and stationary Laplace distribution.

\subsection{Brownian motion with a delta killing point, without resetting}
\label{sec:noreset-kill}

Before combining resetting and killing, it is useful to solve the elementary problem of diffusion with a localized killing term but without resetting. The non-normalized propagator $q_k(x,t|x_0)$ satisfies
\begin{equation}
    \partial_t q_k(x,t|x_0)
    =D\partial_x^2 q_k(x,t|x_0)-k\delta(x)q_k(0,t|x_0),
    \qquad
    q_k(x,0|x_0)=\delta(x-x_0).
    \label{eq:qPDE}
\end{equation}
Equivalently, the Feynman--Kac weight is
\begin{equation}
    \exp\left[-k\int_0^t\delta(X_u)\,\mathrm{d} u\right],
\end{equation}
so that the killing is partial and local. It is not the same as imposing an absorbing boundary at $x=0$, except in the limit $k\to\infty$.

The Laplace transform obeys
\begin{equation}
    \left[s-D\partial_x^2+k\delta(x)\right]
    \widetilde q_k(x,s|x_0)=\delta(x-x_0).
\end{equation}
Since the delta killing term is a rank-one perturbation, the full resolvent is given by the rank-one resolvent formula \cite{SimonRankOne,AlbeverioPointInteractions}, equivalently obtained in the physics literature by an exact resummation of the Dyson series for repeated interactions with the point trap \cite{MazzoloMonthusKilling}:
\begin{equation}
    \widetilde q_k(x,s|x_0)
    =\widetilde G_0(x,s|x_0)
    -\frac{k\widetilde G_0(x,s|0)\widetilde G_0(0,s|x_0)}
          {1+k\widetilde G_0(0,s|0)} .
\end{equation}
With $\lambda=\sqrt{s/D}$, the exact killed resolvent becomes
\begin{equation}
    \widetilde q_k(x,s|x_0)
    =\frac{1}{2D\lambda}
    \left[
        \mathrm{e}^{-\lambda|x-x_0|}
        -\frac{k}{2D\lambda+k}\,
         \mathrm{e}^{-\lambda(|x|+|x_0|)}
    \right].
    \label{eq:qLaplace}
\end{equation}
The continuity of $q_k$ at the origin and integration of Eq.~\eqref{eq:qPDE} through $x=0$ imply the jump condition
\begin{equation}
    D\left[\partial_x q_k(0^+,t|x_0)-\partial_x q_k(0^-,t|x_0)\right]
    =kq_k(0,t|x_0).
\end{equation}

The inverse Laplace transform of Eq.~\eqref{eq:qLaplace} follows from the standard inversion formula in Ref.~\cite{Carslaw} (p.~495, Eq.~(13)):
\begin{equation}
    \begin{aligned}
    q_k(x,t|x_0)
    =&\;G_0(x,t|x_0)
    \\
    &-\frac{k}{4D}
    \exp\left[
       \frac{k(|x|+|x_0|)}{2D}+\frac{k^2t}{4D}
    \right]
    \operatorname{erfc}\left(
       \frac{|x|+|x_0|}{2\sqrt{Dt}}+\frac{k\sqrt t}{2\sqrt D}
    \right),
    \end{aligned}
    \label{eq:qTime}
\end{equation}
This result also appears in Ref.~\cite{MazzoloMonthusKilling}. It has the correct limiting cases: for $k=0$ it reduces to the free Gaussian propagator, while for $k\to\infty$ it reduces to the image solution for a perfectly absorbing point at the origin~\cite{Redner}.

The survival probability without resetting is
\begin{equation}
    Q_k(|x_0|,t)=\int_{-\infty}^{\infty}q_k(x,t|x_0)\,\mathrm{d} x.
\end{equation}
Because the only loss of probability is through the delta killing point,
\begin{equation}
    \frac{\mathrm{d} Q_k}{\mathrm{d} t}=-kq_k(0,t|x_0).
\end{equation}
In Laplace space,
\begin{equation}
    s\widetilde Q_k(|x_0|,s)-1=-k\widetilde q_k(0,s|x_0),
\end{equation}
which yields the exact survival transform
\begin{equation}
    \widetilde Q_k(|x_0|,s)
    =\frac{1}{s}
    \left[
        1-\frac{k\mathrm{e}^{-|x_0|\sqrt{s/D}}}{2\sqrt{Ds}+k}
    \right].
    \label{eq:QkLaplace}
\end{equation}
The corresponding time-domain survival probability is
\begin{equation}
    Q_k(|x_0|,t)
    =\operatorname{erf}\left(\frac{|x_0|}{2\sqrt{Dt}}\right)
    +\exp\left[\frac{k|x_0|}{2D}+\frac{k^2t}{4D}\right]
     \operatorname{erfc}\left(
        \frac{|x_0|}{2\sqrt{Dt}}+\frac{k\sqrt t}{2\sqrt D}
     \right).
    \label{eq:QkTime}
\end{equation}
This expression was previously derived in Ref.~\cite{SzaboLammWeiss1984}. For $k\to\infty$, it becomes $Q_\infty(|x_0|,t)=\operatorname{erf}[|x_0|/(2\sqrt{Dt})]$, the usual survival probability before first hitting the origin~\cite{Redner}.

\subsection{Propagator with stochastic resetting and delta killing}
\label{sec:reset-kill}
We now combine Poissonian resetting to $x_0$ at rate $r$ with the
delta killing point $k\delta(x)$. Let $P(x,t|x_0)$ be the
non-normalized density of particles that are still alive at time
$t$. Its total mass is the survival probability
\begin{equation}
    S(t)=\int_{-\infty}^{\infty}P(x,t|x_0)\,\mathrm{d} x.
\end{equation}
The forward equation is
\begin{equation}
    \partial_tP(x,t|x_0)
    =D\partial_x^2P(x,t|x_0)-rP(x,t|x_0)
    +rS(t)\delta(x-x_0)-kP(0,t|x_0)\delta(x).
    \label{eq:FullPDE}
\end{equation}
The factor $S(t)$ in the resetting source is important: only
particles that have not yet been killed can reset. Indeed, the total
mass removed from all positions by resetting at time $t$ is
$r\int P(x,t|x_0)\,\mathrm{d} x=rS(t)$, and this mass is entirely
redeposited at $x_0$; the source term must therefore scale with the
instantaneous surviving mass $S(t)$ rather than being fixed.

We also note a point concerning the forward equation of Whitehouse,
Evans, and Majumdar~\cite{Whitehouse2013} [their Eq.~(1)], whose
resetting source is instead written as a fixed rate
$r\delta(x-x_0)$, independent of $S(t)$. This form is exact for the
killing-free process, where the surviving mass is identically $1$ at
all times [cf.\ Eq.~\eqref{eq:resetMaster}]; once $k>0$, however,
mass conservation of the resetting mechanism requires the
$S(t)$-dependent source used above. This point does not affect the
results of Ref.~\cite{Whitehouse2013}, since their derivation
proceeds entirely from the backward equation [their Eq.~(2)], where
the resetting generator appears in the standard adjoint form
$r[q(x_0,t)-q(x,t)]$ and the issue does not arise.
%The results of Ref.~\cite{Whitehouse2013} are nevertheless unaffected, since they are derived from the backward equation, where the resetting generator appears in the standard form $r[q(x_0,t)-q(x,t)]$.

Integrating Eq.~\eqref{eq:FullPDE} over $x$ gives the exact mass balance
\begin{equation}
    \frac{\mathrm{d} S(t)}{\mathrm{d} t}=-kP(0,t|x_0).
    \label{eq:massbalance}
\end{equation}
Note that the resetting conserves the mass of surviving trajectories; the delta killing point is the only mechanism that decreases $S(t)$.

\subsubsection{Renewal representation}

The process renews at each reset, but renewal is now conditioned on survival up to the reset time. Split trajectories according to the time $\tau$ elapsed since the last reset.

First, if there has been no reset up to time $t$, the contribution is
\begin{equation}
    \mathrm{e}^{-rt}q_k(x,t|x_0).
\end{equation}
Second, if the last reset occurred at time $t-\tau$, the particle must have survived the full resetting--killing dynamics during the interval $[0,t-\tau]$, which has probability $S(t-\tau)$; it then resets with rate $r$, undergoes no further reset during the last interval of duration $\tau$, and diffuses with the delta killing point. This gives
\begin{equation}
    P(x,t|x_0)
    =\mathrm{e}^{-rt}q_k(x,t|x_0)
    +r\int_0^t \mathrm{e}^{-r\tau}q_k(x,\tau|x_0)S(t-\tau)\,\mathrm{d}\tau .
    \label{eq:Prenewal}
\end{equation}
Integrating Eq.~\eqref{eq:Prenewal} over $x$ gives the renewal equation for the survival probability:
\begin{equation}
    S(t)
    =\mathrm{e}^{-rt}Q_k(|x_0|,t)
    +r\int_0^t \mathrm{e}^{-r\tau}Q_k(|x_0|,\tau)S(t-\tau)\,\mathrm{d}\tau .
    \label{eq:Srenewal}
\end{equation}
Equations~\eqref{eq:Prenewal} and \eqref{eq:Srenewal} are exact time-domain formulas. They are often the most transparent representation: the elementary building block is the killed Brownian propagator without resets, while the full effect of resetting enters through a renewal convolution.

\subsubsection{Laplace-space solution}

Taking the Laplace transform of Eq.~\eqref{eq:Prenewal} gives
\begin{equation}
    \widetilde P(x,s|x_0)
    =\widetilde q_k(x,s+r|x_0)\left[1+r\widetilde S(s)\right].
    \label{eq:PLapPre}
\end{equation}
Similarly, Eq.~\eqref{eq:Srenewal} gives
\begin{equation}
    \widetilde S(s)
    =\widetilde Q_k(|x_0|,s+r)\left[1+r\widetilde S(s)\right].
\end{equation}
Therefore
\begin{equation}
    \widetilde S(s)
    =\frac{\widetilde Q_k(|x_0|,s+r)}{1-r\widetilde Q_k(|x_0|,s+r)},
    \qquad
    \widetilde P(x,s|x_0)
    =\frac{\widetilde q_k(x,s+r|x_0)}{1-r\widetilde Q_k(|x_0|,s+r)}.
    \label{eq:generalLaplaceRenewal}
\end{equation}
This is a compact and useful exact solution.

It is convenient to introduce
\begin{equation}
    u=s+r,
    \qquad
    \lambda=\sqrt{\frac{s+r}{D}},
\end{equation}
and
\begin{equation}
    B(s)=\frac{k\mathrm{e}^{-|x_0|\lambda}}{2D\lambda+k}
    =\frac{k\exp[-|x_0|\sqrt{(s+r)/D}]}
           {2\sqrt{D(s+r)}+k}.
    \label{eq:Bdef}
\end{equation}
Using Eq.~\eqref{eq:QkLaplace},
\begin{equation}
    \widetilde Q_k(|x_0|,s+r)=\frac{1-B(s)}{s+r}.
\end{equation}
Hence
\begin{equation}
    1-r\widetilde Q_k(|x_0|,s+r)
    =\frac{s+rB(s)}{s+r}.
\end{equation}
The survival probability therefore has the explicit Laplace transform
\begin{equation}
    \widetilde S(s)
    =\frac{1-B(s)}{s+rB(s)}.
    \label{eq:SLaplaceFinal}
\end{equation}
Likewise, using Eq.~\eqref{eq:qLaplace} at $s+r$, the non-normalized propagator is
\begin{equation}
    \begin{aligned}
    \widetilde P(x,s|x_0)
    =&\;\frac{s+r}{s+rB(s)}\,
      \frac{1}{2D\lambda}
      \left[
        \mathrm{e}^{-\lambda|x-x_0|}
        -\frac{k}{2D\lambda+k}\,
         \mathrm{e}^{-\lambda(|x|+|x_0|)}
      \right],
    \\
    &\lambda=\sqrt{\frac{s+r}{D}}.
    \end{aligned}
    \label{eq:PLaplaceFinal}
\end{equation}
Equations~\eqref{eq:SLaplaceFinal} and \eqref{eq:PLaplaceFinal} are the main exact solution for resetting plus delta killing.

\subsubsection{Checks and limiting cases}

The formulas above have several useful limits.

\paragraph*{No killing, $k=0$.}
In this case $B(s)=0$, so
\begin{equation}
    \widetilde S(s)=\frac{1}{s},
\end{equation}
as expected for a normalized process without absorption. Equation~\eqref{eq:PLaplaceFinal} reduces to Eq.~\eqref{eq:resetLaplace}, the standard resetting propagator~\cite{EvansMajumdar2011,EvansMajumdarSchehr2020}.

\paragraph*{No resetting, $r=0$.}
Equation~\eqref{eq:generalLaplaceRenewal} reduces to
\begin{equation}
    \widetilde S(s)=\widetilde Q_k(|x_0|,s),
    \qquad
    \widetilde P(x,s|x_0)=\widetilde q_k(x,s|x_0),
\end{equation}
which is precisely the killed Brownian motion of Sec.~\ref{sec:noreset-kill}.

\paragraph*{Perfectly absorbing target, $k\to\infty$.}
In this limit
\begin{equation}
    B(s)\longrightarrow \exp\left[-|x_0|\sqrt{\frac{s+r}{D}}\right].
\end{equation}
Therefore
\begin{equation}
    \widetilde S_\infty(s)
    =\frac{1-\exp[-|x_0|\sqrt{(s+r)/D}]}
           {s+r\exp[-|x_0|\sqrt{(s+r)/D}]}.
    \label{eq:PerfectAbsResetSurvival}
\end{equation}
This is the classical Laplace-space survival probability for a Brownian searcher reset to $x_0$ at rate $r$ and absorbed when it first reaches the target at the origin~\cite{EvansMajumdar2011,EvansMajumdarSchehr2020}. Thus the finite-$k$ delta killing point is a partial-absorption generalization of the standard absorbing-target problem.

\paragraph*{Mean absorption time.}
The mean lifetime is
\begin{equation}
    T=\int_0^\infty S(t)\,\mathrm{d} t=\widetilde S(0).
\end{equation}
From Eq.~\eqref{eq:SLaplaceFinal},
\begin{equation}
    T
    =\frac{1-B(0)}{rB(0)}
    =\frac{1}{r}
      \left[
        \left(1+\frac{2\sqrt{Dr}}{k}\right)
        \exp\left(|x_0|\sqrt{\frac{r}{D}}\right)-1
      \right].
    \label{eq:MeanLifetime}
\end{equation}
This exact formula was first derived by Whitehouse, Evans, and
Majumdar~\cite{Whitehouse2013}, Eq.~(36) therein, using the
backward-equation formulation. We recover it here from the
forward-equation renewal construction of Sec.~\ref{sec:reset-kill},
for consistency with the treatment of drift in Sec.~III.

For $k\to\infty$ this reduces to the perfect-target result derived in~\cite{EvansMajumdar2011,EvansMajumdarSchehr2020}
\begin{equation}
    T_\infty=\frac{1}{r}\left[
        \exp\left(|x_0|\sqrt{\frac{r}{D}}\right)-1
    \right]
\end{equation}
for a perfectly absorbing target with stochastic resetting.

\subsection{Long-time behavior}
\label{sec:longtime}

The large-time behavior is governed by the singularity of $\widetilde S(s)$, or equivalently of $\widetilde P(x,s|x_0)$, with the largest real part in the complex $s$ plane. The branch point associated with the square root in $B(s)$ is at
\begin{equation}
    s=-r.
\end{equation}
For $r>0$ and $k>0$, however, there is a simple pole at $s=-\theta$ with
\begin{equation}
    0<\theta<r,
\end{equation}
which dominates the branch-cut contribution. Thus the survival probability decays exponentially, not algebraically.

\subsubsection{Dominant pole and decay rate}

Write
\begin{equation}
    \widetilde S(s)=\frac{N(s)}{F(s)},
    \qquad
    N(s)=1-B(s),
    \qquad
    F(s)=s+rB(s).
\end{equation}
A pole occurs when
\begin{equation}
    F(s)=0.
\end{equation}
Setting $s=-\theta$ and defining
\begin{equation}
    \lambda_*=\sqrt{\frac{r-\theta}{D}},
\end{equation}
one obtains
\begin{equation}
    -\theta+r\frac{k\mathrm{e}^{-|x_0|\lambda_*}}{2D\lambda_*+k}=0.
\end{equation}
The central long-time result is therefore the following implicit equation for the asymptotic decay rate:
\begin{equation}
    \theta
    =r\frac{k\exp[-|x_0|\sqrt{(r-\theta)/D}]}
            {2\sqrt{D(r-\theta)}+k}.
    \label{eq:thetaEquation}
\end{equation}
This scalar equation coincides with the pole equation obtained
directly by Whitehouse, Evans, and Majumdar~\cite{Whitehouse2013},
their Eq.~(28), for a diffusive searcher with resetting and a
finite-reactivity target. It is also the finite-reactivity analogue of the standard absorbing-target equation~\cite{EvansMajumdar2011,EvansMajumdarSchehr2020}. It can be written in dimensionless form. Let
\begin{equation}
    z=|x_0|\sqrt{\frac{r}{D}},
    \qquad
    \kappa=\frac{k}{2\sqrt{Dr}},
    \qquad
    \phi=\frac{\theta}{r}.
\end{equation}
Then Eq.~\eqref{eq:thetaEquation} becomes
\begin{equation}
    \phi=\frac{\kappa\exp[-z\sqrt{1-\phi}]}
               {\sqrt{1-\phi}+\kappa},
    \qquad 0<\phi<1.
    \label{eq:thetaDimensionless}
\end{equation}
The branch point is at $s=-r$, while the pole is at $s=-\theta$ with $\theta<r$; the pole is therefore closer to the origin and controls the leading long-time decay.

In the perfectly absorbing limit $k\to\infty$, Eq.~\eqref{eq:thetaEquation} reduces to
\begin{equation}
    \theta_\infty
    =r\exp\left[-|x_0|\sqrt{\frac{r-\theta_\infty}{D}}\right],
    \label{eq:thetaPerfect}
\end{equation}
which is the pole equation found in the standard resetting literature for a perfectly absorbing target~\cite{EvansMajumdar2011,EvansMajumdarSchehr2020}. The finite-$k$ delta killing point multiplies the perfect-absorption factor by the reactivity factor
\begin{equation}
    \frac{k}{2\sqrt{D(r-\theta)}+k}.
\end{equation}
This factor lies between $0$ and $1$ and decreases the absorption rate relative to the perfect target.

\subsubsection{Residue and survival amplitude}

The leading asymptotic survival probability has the form
\begin{equation}
    S(t)\simeq A\mathrm{e}^{-\theta t},
    \qquad t\to\infty,
    \label{eq:Sasymptotic}
\end{equation}
where the amplitude is the residue of $\widetilde S(s)$ at $s=-\theta$:
\begin{equation}
    A=\frac{N(-\theta)}{F'(-\theta)}.
\end{equation}
Since $B(-\theta)=\theta/r$, one has
\begin{equation}
    N(-\theta)=1-\frac{\theta}{r}.
\end{equation}
Moreover,
\begin{equation}
    B'(s)\big|_{s=-\theta}
    =-\frac{\theta}{r}
      \left[
        \frac{|x_0|}{2D\lambda_*}
        +\frac{1}{\lambda_*(2D\lambda_*+k)}
      \right].
\end{equation}
Therefore
\begin{equation}
    F'(-\theta)
    =1+rB'(-\theta)
    =1-\theta
      \left[
        \frac{|x_0|}{2D\lambda_*}
        +\frac{1}{\lambda_*(2D\lambda_*+k)}
      \right].
\end{equation}
The amplitude is thus
\begin{equation}
    A=\frac{1-\theta/r}
           {1-\theta\left[
              \dfrac{|x_0|}{2D\lambda_*}
              +\dfrac{1}{\lambda_*(2D\lambda_*+k)}
           \right]}.
    \label{eq:Amplitude}
\end{equation}
The branch-cut correction from $s=-r$ is exponentially smaller, of order $\mathrm{e}^{-rt}$ times an algebraic power, and is subdominant because $\theta<r$.

\subsubsection{Long-time propagator and quasi-stationary profile}

The full propagator has the same dominant pole. From Eq.~\eqref{eq:PLaplaceFinal}, the residue of $\widetilde P(x,s|x_0)$ at $s=-\theta$ is
\begin{equation}
    \frac{(r-\theta)\widetilde q_k(x,r-\theta|x_0)}{F'(-\theta)}.
\end{equation}
Combining this with the survival residue in Eq.~\eqref{eq:Amplitude}, one obtains
\begin{equation}
    P(x,t|x_0)\simeq A\,p_\mathrm{qs}(x)\,\mathrm{e}^{-\theta t},
    \qquad t\to\infty,
    \label{eq:Pasymptotic}
\end{equation}
where the normalized quasi-stationary profile is
\begin{equation}
    p_\mathrm{qs}(x)=r\widetilde q_k(x,r-\theta|x_0).
    \label{eq:pqsCompact}
\end{equation}
Using Eq.~\eqref{eq:qLaplace}, this is explicitly
\begin{equation}
    p_\mathrm{qs}(x)
    =\frac{r}{2D\lambda_*}
      \left[
        \mathrm{e}^{-\lambda_*|x-x_0|}
        -\frac{k}{2D\lambda_*+k}\,
         \mathrm{e}^{-\lambda_*(|x|+|x_0|)}
      \right].
    \label{eq:pqsExplicit}
\end{equation}
The normalization follows directly from the pole equation:
\begin{equation}
    \int_{-\infty}^{\infty}p_\mathrm{qs}(x)\,\mathrm{d} x
    =r\widetilde Q_k(|x_0|,r-\theta)
    =r\frac{1-B(-\theta)}{r-\theta}
    =1.
\end{equation}
Thus, conditioned on survival, the spatial distribution converges to $p_\mathrm{qs}(x)$.

\subsubsection{Weak and strong killing regimes}

For weak reactivity, $k\ll 2\sqrt{Dr}$, one may solve Eq.~\eqref{eq:thetaEquation} perturbatively by replacing $\theta$ by $0$ on the right-hand side. This gives
\begin{equation}
    \theta
    =\frac{k}{2}\sqrt{\frac{r}{D}}
      \exp\left[-|x_0|\sqrt{\frac{r}{D}}\right]
      +O(k^2).
    \label{eq:WeakKillingTheta}
\end{equation}
The leading term has a simple interpretation:
\begin{equation}
    \theta\simeq k p_r^\mathrm{st}(0|x_0),
\end{equation}
where $p_r^\mathrm{st}$ is the stationary density of the resetting process without killing, Eq.~\eqref{eq:stationaryReset}. Thus, for weak delta killing, the decay rate is the killing strength times the stationary probability density at the delta killing point.

For strong reactivity, $k\gg 2\sqrt{D(r-\theta)}$, Eq.~\eqref{eq:thetaEquation} approaches the perfect-target equation~\eqref{eq:thetaPerfect}. To isolate the finite-reactivity correction, rewrite the factor multiplying the perfect-absorption contribution as
\begin{equation}
\frac{k}{2\sqrt{D(r-\theta)}+k}
=
\frac{1}{
1+\dfrac{2\sqrt{D(r-\theta)}}{k}
}  =
1-\frac{2\sqrt{D(r-\theta)}}{k}
+O(k^{-2}).
\label{eq:strong-k-expansion}
\end{equation}
Thus, at fixed $\theta$, the departure from perfect absorption is of relative order $k^{-1}$. Since this factor is smaller than unity for every finite $k$, finite reactivity decreases the right-hand side of Eq.~\eqref{eq:thetaEquation} and consequently lowers the asymptotic decay rate relative to that of a perfectly absorbing target.

The correction to the pole itself can also be obtained explicitly. Let $\theta_\infty$ denote the physical solution of the perfect-target equation~\eqref{eq:thetaPerfect}, and write
\begin{equation}
\theta_k
=
\theta_\infty+\frac{c_1}{k}+O(k^{-2}).
\end{equation}
Expanding Eq.~\eqref{eq:thetaEquation} to first order in $k^{-1}$ yields
\begin{equation}
c_1
=
-
\frac{
2\theta_\infty
\sqrt{D(r-\theta_\infty)}
}{
1-
\dfrac{|x_0|\theta_\infty}{
2\sqrt{D(r-\theta_\infty)}
}
}.
\end{equation}
Hence,
\begin{equation}
\theta_k
=
\theta_\infty
-
\frac{
2\theta_\infty\sqrt{D(r-\theta_\infty)}
}{
k\left[
1-
\dfrac{|x_0|\theta_\infty}{
2\sqrt{D(r-\theta_\infty)}
}
\right]
}
+O(k^{-2}).
\label{eq:strong-k-theta-correction}
\end{equation}
For the physical pole below the branch point, the denominator is positive. The leading correction is therefore negative, confirming that an imperfect delta killing trap produces a smaller decay rate than a perfectly absorbing target. The perfect-target result is recovered continuously as $k\to\infty$.

\subsubsection{Comparison with the no-reset case}

When $r=0$, the survival is the no-reset result $Q_k(|x_0|,t)$ of Eq.~\eqref{eq:QkTime}. Its large-time behavior follows from the small-$s$ expansion of Eq.~\eqref{eq:QkLaplace}:
\begin{equation}
    \widetilde Q_k(|x_0|,s)
    =\frac{|x_0|+2D/k}{\sqrt{D}}\,s^{-1/2}+O(1),
    \qquad s\to 0.
\end{equation}
Since $Q_k(|x_0|,t)$ is nonnegative and nonincreasing, the
%Hardy--Littlewood--Karamata
Tauberian theorem for Laplace transforms, together with the monotone density theorem \cite{FellerTauberian}, implies that
\begin{equation}
\widetilde Q_k(|x_0|,s)\sim C s^{-\rho}
\quad\Longrightarrow\quad
Q_k(|x_0|,t)\sim\frac{C}{\Gamma(\rho)}t^{\rho-1}.
\end{equation}
Taking $\rho=1/2$ and
$C=(|x_0|+2D/k)/\sqrt{D}$ therefore yields
\begin{equation}
    Q_k(|x_0|,t)
    \simeq
    \frac{|x_0|+2D/k}{\sqrt{\pi D\,t}},
    \qquad t\to\infty,
    \qquad r=0.
    \label{eq:NoResetAsymptotic}
\end{equation}

Thus the localized killing term eventually kills the particle with probability one, but the survival is only algebraic. Resetting changes this algebraic decay into the exponential law \eqref{eq:Sasymptotic}. The resetting mechanism repeatedly brings the searcher back to $x_0$ and produces statistically independent attempts to reach and react at the delta killing point; this renewal structure is what moves the dominant singularity from the branch point at $s=0$ in the no-reset case to a pole at $s=-\theta$ in the resetting case.

%\subsection{Summary}

%For a Brownian particle reset to $x_0$ at rate $r$ and partially absorbed by a delta killing point $k\delta(x)$ at the origin, the exact non-normalized propagator and survival probability are most compactly written in Laplace space. Defining
%\begin{equation}
%    B(s)=\frac{k\exp[-|x_0|\sqrt{(s+r)/D}]}
%           {2\sqrt{D(s+r)}+k},
%\end{equation}
%one has
%\begin{equation}
%    \widetilde S(s)=\frac{1-B(s)}{s+rB(s)}
%\end{equation}
%and
%\begin{equation}
%    \widetilde P(x,s|x_0)
%    =\frac{s+r}{s+rB(s)}\,
%      \frac{1}{2D\lambda}
%      \left[
%        \mathrm{e}^{-\lambda|x-x_0|}
%        -\frac{k}{2D\lambda+k}\,
%         \mathrm{e}^{-\lambda(|x|+|x_0|)}
%      \right],
%    \qquad
%    \lambda=\sqrt{\frac{s+r}{D}}.
%\end{equation}
%The standard resetting propagator is recovered at $k=0$, the killed Brownian propagator is recovered at $r=0$, and the usual absorbing-target resetting problem is recovered at $k\to\infty$~\cite{EvansMajumdar2011,EvansMajumdarSchehr2020}. For $r>0$ and $k>0$, the long-time survival is
%\begin{equation}
%    S(t)\simeq A\mathrm{e}^{-\theta t},
%\end{equation}
%where $\theta$ is determined by Eq.~\eqref{eq:thetaEquation}. Without resetting, the survival instead decays as $t^{-1/2}$.

\section{Brownian Motion with Constant Drift, Stochastic Resetting, and a Delta Killing Point}
We now extend the model of Sec.~II by adding a constant drift $\mu\in\mathbb R$; all other parameters and conventions retain the meanings introduced above. We first analyze the drifted dynamics with resetting and with delta killing separately, and then combine the two mechanisms through the same renewal structure as in the driftless case. This organization separates the effect of drift-induced transience from the renewal effect of resetting. We study the stochastic process
\begin{equation}
    \mathrm{d}X_t=\mu\,\mathrm{d}t+\sqrt{2D}\,\mathrm{d}W_t,
    \label{eq:SDEdrift}
\end{equation}
where $\mu$ is the newly introduced constant drift. The free forward generator is
\begin{equation}
    \mathcal L = D\partial_x^2-\mu\partial_x.
\end{equation}
The free drifted propagator is
\begin{equation}
    G_\mu(x,t|x_0)
    =\frac{1}{\sqrt{4\pi Dt}}
    \exp\left[-\frac{(x-x_0-\mu t)^2}{4Dt}\right].
    \label{eq:freeDriftTime}
\end{equation}
Its Laplace transform is most compactly written in terms of
\begin{equation}
    \Gamma(s)=\sqrt{\mu^2+4Ds},
    \qquad \operatorname{Re}\Gamma(s)>0.
    \label{eq:GammaDefinition}
\end{equation}
The free drifted resolvent is therefore
\begin{equation}
    \widetilde G_\mu(x,s|y)
    =\int_0^\infty \mathrm{e}^{-st}G_\mu(x,t|y)\,\mathrm{d}t
    =\frac{1}{\Gamma(s)}
    \exp\left[\frac{\mu(x-y)-\Gamma(s)|x-y|}{2D}\right].
    \label{eq:GmuLaplace}
\end{equation}
The driftless formula is recovered by taking $\mu=0$, since then
\begin{equation}
    \Gamma(s)=2\sqrt{Ds}.
\end{equation}
The reset point remains the initial point $x_0$, and the delta killing point remains at the origin. The sign of $\mu x_0$ is important: if $\mu x_0>0$, the drift pushes the particle away from the killing point, whereas if $\mu x_0<0$, it pushes it toward the killing point.

\subsection{Drifted Brownian motion with resetting, without killing}
\label{sec:reset-no-kill-drift}

Let $p_r(x,t|x_0)$ be the normalized propagator of a drifted Brownian particle which resets to $x_0$ at Poisson rate $r$. It satisfies
\begin{equation}
    \partial_t p_r(x,t|x_0)
    =D\partial_x^2p_r(x,t|x_0)-\mu\partial_xp_r(x,t|x_0)
    -rp_r(x,t|x_0)+r\delta(x-x_0),
    \label{eq:resetDriftMaster}
\end{equation}
with $p_r(x,0|x_0)=\delta(x-x_0)$. The term $-rp_r$ removes probability density from all positions due to resets, while the source $r\delta(x-x_0)$ reinjects the same total probability at the reset point.

A renewal decomposition over the time $\tau$ elapsed since the last reset gives
\begin{equation}
    p_r(x,t|x_0)
    =\mathrm{e}^{-rt}G_\mu(x,t|x_0)
    +r\int_0^t \mathrm{e}^{-r\tau}G_\mu(x,\tau|x_0)\,\mathrm{d}\tau.
    \label{eq:resetDriftRenewal}
\end{equation}
The first term corresponds to trajectories with no reset before time $t$; the second term corresponds to trajectories whose last reset occurred at time $t-\tau$.

In Laplace space,
\begin{equation}
    \widetilde p_r(x,s|x_0)
    =\left(1+\frac{r}{s}\right)\widetilde G_\mu(x,s+r|x_0).
\end{equation}
Using Eq.~\eqref{eq:GmuLaplace}, one obtains
\begin{equation}
    \widetilde p_r(x,s|x_0)
    =\frac{s+r}{s}\,
     \frac{1}{\Gamma(s+r)}
     \exp\left[\frac{\mu(x-x_0)-\Gamma(s+r)|x-x_0|}{2D}\right].
    \label{eq:resetDriftLaplace}
\end{equation}
The stationary state is obtained either by taking the residue at the pole $s=0$ or by integrating over the last reset age:
\begin{equation}
    p_r^{\mathrm{st}}(x|x_0)
    =r\widetilde G_\mu(x,r|x_0).
\end{equation}
Thus
\begin{equation}
    p_r^{\mathrm{st}}(x|x_0)
    =\frac{r}{\Gamma(r)}
     \exp\left[\frac{\mu(x-x_0)-\Gamma(r)|x-x_0|}{2D}\right],
    \qquad
    \Gamma(r)=\sqrt{\mu^2+4Dr}.
    \label{eq:stationaryResetDrift}
\end{equation}
This distribution is asymmetric when $\mu\ne0$. For $x>x_0$ the exponential rate is $[\Gamma(r)-\mu]/(2D)$, while for $x<x_0$ it is $[\Gamma(r)+\mu]/(2D)$. The two rates are positive because $\Gamma(r)>|\mu|$ for $r>0$. Therefore resetting confines the process and produces a normalized nonequilibrium steady state even in the presence of a constant drift.

For $\mu=0$, Eq.~\eqref{eq:stationaryResetDrift} reduces to the usual Laplace distribution
\begin{equation}
    p_r^{\mathrm{st}}(x|x_0)
    =\frac{1}{2}\sqrt{\frac{r}{D}}\,
    \exp\left[-\sqrt{\frac{r}{D}}|x-x_0|\right],
\end{equation}
which is the standard result in the resetting literature \cite{EvansMajumdar2011,EvansMajumdarSchehr2020}.

\subsection{Drifted Brownian motion with a delta killing point, without resetting}
\label{sec:drift-kill-no-reset}

Before combining resetting and killing, we solve the drifted diffusion problem with a localized killing term only. To distinguish this process from the driftless one of Sec.~\ref{sec:noreset-kill}, we denote its propagator and survival probability by $q_{k,\mu}$ and $Q_{k,\mu}$, respectively. Let $q_{k,\mu}(x,t|x_0)$ denote the non-normalized propagator. It satisfies
\begin{equation}
    \partial_t q_{k,\mu}(x,t|x_0)
    =D\partial_x^2q_{k,\mu}(x,t|x_0)-\mu\partial_xq_{k,\mu}(x,t|x_0)
    -k\delta(x)q_{k,\mu}(0,t|x_0),
    \label{eq:qDriftPDE}
\end{equation}
with $q_{k,\mu}(x,0|x_0)=\delta(x-x_0)$. The Feynman--Kac interpretation is
\begin{equation}
    q_{k,\mu}(x,t|x_0)\,\mathrm{d}x
    =\mathbb P_{x_0}\left[X_t\in\mathrm{d}x\,;
     \exp\left(-k\int_0^t\delta(X_u)\,\mathrm{d}u\right)\right],
\end{equation}
so the killing is partial and local. The quantity $L_t^0=\int_0^t\delta(X_u)\,\mathrm{d}u$ is the local time of the process at the origin, i.e., the density, with respect to Lebesgue measure, of the time the trajectory spends in an infinitesimal neighborhood of $x=0$ up to time $t$~\cite{BorodinSalminen}. Survival to time $t$ therefore occurs with probability $\exp(-kL_t^0)$: the particle is killed by an independent exponential clock of rate $k$ run in the local time accumulated at the origin, rather than instantaneously upon first arrival. This is different from an absorbing boundary, except in the limit $k\to\infty$.

The Laplace transform solves
\begin{equation}
    \left[s-D\partial_x^2+\mu\partial_x+k\delta(x)\right]
    \widetilde q_{k,\mu}(x,s|x_0)=\delta(x-x_0).
    \label{eq:qDriftResolventEquation}
\end{equation}
The delta perturbation is rank one, hence the resolvent identity \cite{SimonRankOne,AlbeverioPointInteractions} gives
\begin{equation}
    \widetilde q_{k,\mu}(x,s|x_0)
    =\widetilde G_\mu(x,s|x_0)
    -\frac{k\widetilde G_\mu(x,s|0)\widetilde G_\mu(0,s|x_0)}
          {1+k\widetilde G_\mu(0,s|0)}.
    \label{eq:qDriftResolventIdentity}
\end{equation}
Since $\widetilde G_\mu(0,s|0)=1/\Gamma(s)$, Eq.~\eqref{eq:qDriftResolventIdentity} becomes
\begin{equation}
    \begin{aligned}
    \widetilde q_{k,\mu}(x,s|x_0)
    =\frac{1}{\Gamma(s)}
    \bigg\{&
    \exp\left[\frac{\mu(x-x_0)-\Gamma(s)|x-x_0|}{2D}\right]
    \\
    &-\frac{k}{\Gamma(s)+k}
    \exp\left[\frac{\mu(x-x_0)-\Gamma(s)(|x|+|x_0|)}{2D}\right]
    \bigg\}.
    \end{aligned}
    \label{eq:qDriftLaplace}
\end{equation}
The first term is the free drifted resolvent. The second term is the contribution generated by the delta killing point. The drift appears only through the exponential gauge factor and through $\Gamma(s)$.

Equation~\eqref{eq:qDriftPDE} also implies a matching condition at the origin. The density is continuous at $x=0$, while integration through a small interval around the origin gives
\begin{equation}
    D\left[\partial_xq_{k,\mu}(0^+,t|x_0)-\partial_xq_{k,\mu}(0^-,t|x_0)\right]
    =kq_{k,\mu}(0,t|x_0).
    \label{eq:jumpConditionDrift}
\end{equation}
The drift term does not contribute to the jump condition because the density itself is continuous at the origin.

The survival probability without resetting is
\begin{equation}
    Q_{k,\mu}(t)=\int_{-\infty}^{\infty}q_{k,\mu}(x,t|x_0)\,\mathrm{d}x.
\end{equation}
The only loss of probability is through the delta killing point, so
\begin{equation}
    \frac{\mathrm{d}Q_{k,\mu}}{\mathrm{d}t}=-kq_{k,\mu}(0,t|x_0).
    \label{eq:survivalFluxDriftNoReset}
\end{equation}
From Eq.~\eqref{eq:qDriftLaplace},
\begin{equation}
    \widetilde q_{k,\mu}(0,s|x_0)
    =\frac{1}{\Gamma(s)+k}
    \exp\left[-\frac{\mu x_0+\Gamma(s)|x_0|}{2D}\right].
    \label{eq:qOriginDriftLaplace}
\end{equation}
The exact no-reset survival transform is therefore
\begin{equation}
    \widetilde Q_{k,\mu}(s)
    =\frac{1}{s}\left[
    1-\frac{k}{\Gamma(s)+k}
    \exp\left[-\frac{\mu x_0+\Gamma(s)|x_0|}{2D}\right]
    \right].
    \label{eq:QDriftLaplace}
\end{equation}
This expression is one of the main building blocks for the resetting calculation below.

Several checks are immediate. If $k=0$, then $\widetilde Q_{k,\mu}(s)=1/s$ and no killing occurs. If $\mu=0$, then $\Gamma(s)=2\sqrt{Ds}$ and Eq.~\eqref{eq:QDriftLaplace} reduces to
\begin{equation}
    \widetilde Q_{k,\mu}(s)=\frac{1}{s}\left[
    1-\frac{k\exp[-|x_0|\sqrt{s/D}]}{2\sqrt{Ds}+k}
    \right],
\end{equation}
which is the driftless formula. Finally, if $k\to\infty$, one obtains the survival transform for a perfectly absorbing point at the origin,
\begin{equation}
    \widetilde Q_\infty(s)
    =\frac{1}{s}\left[
    1-\exp\left[-\frac{\mu x_0+\Gamma(s)|x_0|}{2D}\right]
    \right].
    \label{eq:QAbsorbingDrift}
\end{equation}
For $x_0>0$ and $\mu>0$, the drift points away from the absorber and the ultimate hitting probability is smaller than one. For $x_0>0$ and $\mu<0$, the drift points toward the absorber and the perfectly absorbing survival tends to zero at long times.

\subsection{Propagator with drift, stochastic resetting, and delta killing}
\label{sec:combined-drift-reset-kill}

We now combine the three ingredients: drift, resetting, and local delta killing. Let $P(x,t|x_0)$ be the non-normalized density of particles that are still alive at time $t$. Its total mass is
\begin{equation}
    S(t)=\int_{-\infty}^{\infty}P(x,t|x_0)\,\mathrm{d}x,
    \label{eq:SurvivalDefinitionCombined}
\end{equation}
which is the survival probability. The forward equation is
\begin{equation}
    \partial_tP(x,t|x_0)
    =D\partial_x^2P(x,t|x_0)-\mu\partial_xP(x,t|x_0)-rP(x,t|x_0)+rS(t)\delta(x-x_0)
    -kP(0,t|x_0)\delta(x).
    \label{eq:combinedForwardDrift}
\end{equation}
As in the driftless case, the reset source is proportional to $S(t)$ rather than to one because only particles that are still alive can reset. Integrating Eq.~\eqref{eq:combinedForwardDrift} over $x$ gives
\begin{equation}
    \frac{\mathrm{d}S}{\mathrm{d}t}=-kP(0,t|x_0).
    \label{eq:combinedFluxRelation}
\end{equation}
The resetting terms conserve the total surviving mass; they merely move surviving particles back to $x_0$. The only loss of mass comes from the delta killing point.

\subsubsection{Renewal representation}

The renewal construction is based on the last reset time. Suppose that the last reset occurred at time $t-\tau$, so that the particle has evolved for a time $\tau$ after this last reset. During this last interval it must not reset again, which gives the factor $\mathrm{e}^{-r\tau}$. It must also survive the delta killing during this interval, and its spatial propagation during this interval is described by the no-reset killed propagator $q_{k,\mu}(x,\tau|x_0)$. The probability that the particle was alive at the time of the last reset is $S(t-\tau)$.

There are two possibilities. Either no reset has occurred at all before time $t$, or the last reset occurred at some time $t-\tau$ with $0<\tau<t$. Hence
\begin{equation}
    P(x,t|x_0)
    =\mathrm{e}^{-rt}q_{k,\mu}(x,t|x_0)
    +r\int_0^t \mathrm{e}^{-r\tau}q_{k,\mu}(x,\tau|x_0)S(t-\tau)\,\mathrm{d}\tau.
    \label{eq:PDriftRenewal}
\end{equation}
The first term is the contribution of trajectories with no reset up to time $t$. The integral term is the contribution of trajectories with a last reset. The factor $r\,\mathrm{d}\tau$ is the probability that a reset occurs in the infinitesimal interval which fixes the last reset age between $\tau$ and $\tau+\mathrm{d}\tau$.

Integrating Eq.~\eqref{eq:PDriftRenewal} over $x$ gives the renewal equation for survival:
\begin{equation}
    S(t)
    =\mathrm{e}^{-rt}Q_{k,\mu}(t)
    +r\int_0^t \mathrm{e}^{-r\tau}Q_{k,\mu}(\tau)S(t-\tau)\,\mathrm{d}\tau.
    \label{eq:SDriftRenewal}
\end{equation}
This equation has a direct probabilistic interpretation. The function $\mathrm{e}^{-rt}Q_{k,\mu}(t)$ is the probability of surviving a reset-free cycle of age $t$. The convolution sums over the age of the last cycle and over the survival of all previous cycles.

\subsubsection{Laplace-space solution}

Taking the Laplace transform of Eq.~\eqref{eq:SDriftRenewal} gives
\begin{equation}
    \widetilde S(s)
    =\widetilde Q_{k,\mu}(s+r)+r\widetilde Q_{k,\mu}(s+r)\widetilde S(s).
\end{equation}
Therefore
\begin{equation}
    \widetilde S(s)
    =\frac{\widetilde Q_{k,\mu}(s+r)}{1-r\widetilde Q_{k,\mu}(s+r)}.
    \label{eq:SDriftLaplaceGeneral}
\end{equation}
Using Eq.~\eqref{eq:QDriftLaplace} at the shifted Laplace variable $s+r$, define
\begin{equation}
    \Gamma_r(s)=\sqrt{\mu^2+4D(s+r)}
    \label{eq:GammaResetShifted}
\end{equation}
and
\begin{equation}
    B_\mu(s)=\frac{k}{\Gamma_r(s)+k}
    \exp\left[-\frac{\mu x_0+\Gamma_r(s)|x_0|}{2D}\right].
    \label{eq:BmuDefinition}
\end{equation}
Then
\begin{equation}
    \widetilde Q_{k,\mu}(s+r)=\frac{1-B_\mu(s)}{s+r}.
\end{equation}
Substitution into Eq.~\eqref{eq:SDriftLaplaceGeneral} yields the first central result of the drifted problem, the exact survival transform
\begin{equation}
    \widetilde S(s)=\frac{1-B_\mu(s)}{s+rB_\mu(s)}.
    \label{eq:SDriftLaplaceFinal}
\end{equation}
This is the survival probability for the full drift--resetting--delta-killing problem.

The same Laplace transform applied to Eq.~\eqref{eq:PDriftRenewal} gives
\begin{equation}
    \widetilde P(x,s|x_0)
    =\widetilde q_{k,\mu}(x,s+r|x_0)
    +r\widetilde q_{k,\mu}(x,s+r|x_0)\widetilde S(s).
\end{equation}
Equivalently,
\begin{equation}
    \widetilde P(x,s|x_0)
    =\frac{\widetilde q_{k,\mu}(x,s+r|x_0)}{1-r\widetilde Q_{k,\mu}(s+r)}.
    \label{eq:PDriftLaplaceGeneral}
\end{equation}
Using again the definition of $B_\mu(s)$, one obtains
\begin{equation}
    \widetilde P(x,s|x_0)
    =\frac{s+r}{s+rB_\mu(s)}\,
    \widetilde q_{k,\mu}(x,s+r|x_0).
    \label{eq:PDriftLaplaceCompact}
\end{equation}
Finally, inserting the explicit expression \eqref{eq:qDriftLaplace}, evaluated at $s+r$, gives
\begin{equation}
    \begin{aligned}
    \widetilde P(x,s|x_0)
    =&\frac{s+r}{s+rB_\mu(s)}\,
    \frac{1}{\Gamma_r(s)}
    \bigg\{
    \exp\left[\frac{\mu(x-x_0)-\Gamma_r(s)|x-x_0|}{2D}\right]
    \\
    &\hspace{2.0cm}
    -\frac{k}{\Gamma_r(s)+k}
    \exp\left[\frac{\mu(x-x_0)-\Gamma_r(s)(|x|+|x_0|)}{2D}\right]
    \bigg\}.
    \end{aligned}
    \label{eq:PDriftLaplaceExplicit}
\end{equation}
Equations~\eqref{eq:SDriftLaplaceFinal} and \eqref{eq:PDriftLaplaceExplicit} are the exact Laplace-space solution of the combined problem.

\begin{figure}[t]
\centering
\includegraphics[width=6in,height=4in]{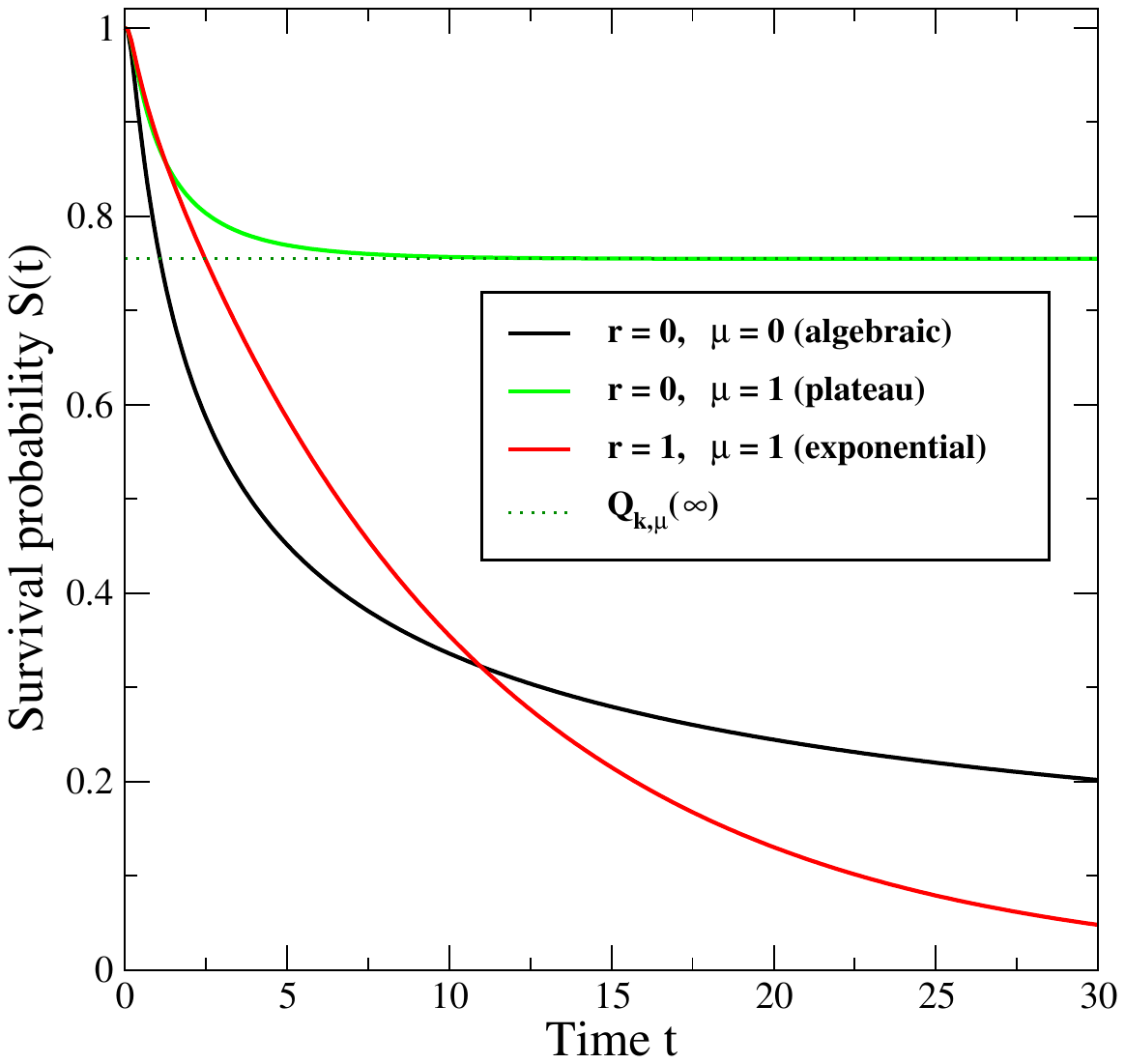}
\caption{Representative survival curves for $D=1$, $k=2$, and $x_0=1$. For $r=0$ and $\mu=0$, the survival decays algebraically to zero. For $r=0$ and $\mu=1$, it approaches the nonzero plateau $Q_{k,\mu}(\infty)$ of Eq.~\eqref{eq:QinftyNoResetDrift}. For $r=1$ and $\mu=1$, resetting removes the plateau and produces exponential decay. The horizontal dashed line is the exact reset-free plateau. The curves were obtained by numerical inversion of the exact Laplace transforms.}
\label{fig:survival-regimes}
\end{figure}

Figure~\ref{fig:survival-regimes} gives a quantitative illustration of the three long-time regimes discussed below. The curves are obtained by numerical inversion of the exact Laplace-space survival formulas derived above.

\paragraph*{Mean absorption time.}
For $r>0$ and $k>0$, the mean absorption time in the presence of drift follows directly by setting $s=0$ in Eq.~\eqref{eq:SDriftLaplaceFinal}:
\begin{equation}
    T_\mu(r)
    =\widetilde S(0)
    =\frac{1}{r}\left[
    \left(1+\frac{\Gamma(r)}{k}\right)
    \exp\left(\frac{\mu x_0+\Gamma(r)|x_0|}{2D}\right)-1
    \right],
    \qquad
    \Gamma(r)=\sqrt{\mu^2+4Dr}.
    \label{eq:MeanLifetimeDrift}
\end{equation}
For $\mu=0$, Eq.~\eqref{eq:MeanLifetimeDrift} exactly recovers the driftless result in Eq.~\eqref{eq:MeanLifetime}. At fixed $|\mu|$, the direction of the drift enters through the factor $\exp(\mu x_0/2D)$, so the optimal resetting rate is not invariant under reversal of the drift. This effect is illustrated in Fig.~\ref{fig:mean-absorption-drift} and recast in dimensionless form in Appendix~\ref{app:optimal-resetting-rate}.

\begin{figure}[t]
\centering
\includegraphics[width=6in,height=4in]{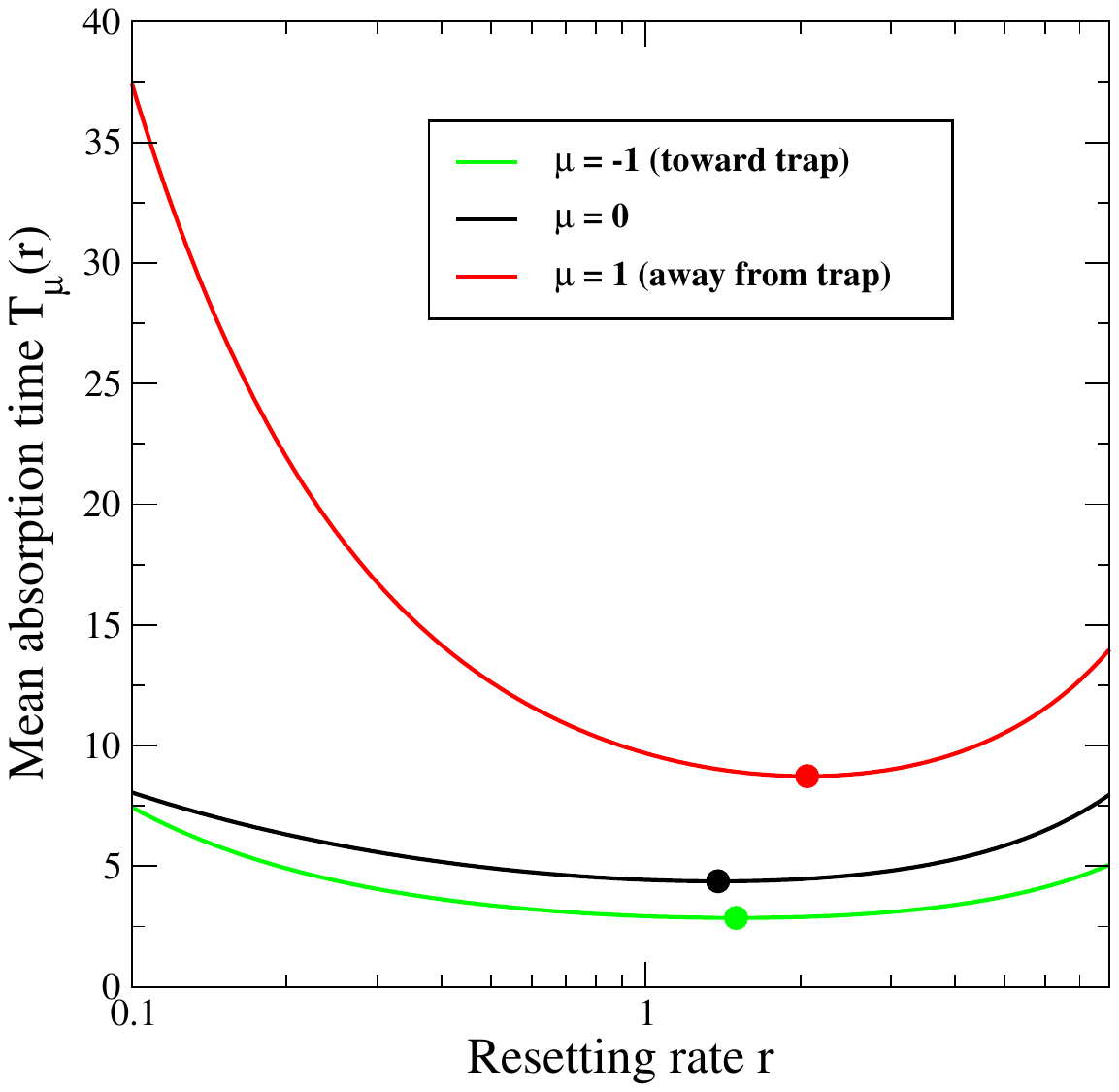}
\caption{Mean absorption time $T_\mu(r)$ from Eq.~\eqref{eq:MeanLifetimeDrift} as a function of the resetting rate for $D=1$, $k=2$, $x_0=1$, and three drift values. The circles mark the numerical minima. For these representative parameters, a drift away from the trap ($\mu=1$) increases the minimum mean time and shifts the optimum to a larger resetting rate than a drift toward the trap ($\mu=-1$).}
\label{fig:mean-absorption-drift}
\end{figure}

\subsubsection{Consistency checks}

Several limits check the result. If $k=0$, then $B_\mu(s)=0$ and $\widetilde S(s)=1/s$. There is no absorption. Equation~\eqref{eq:PDriftLaplaceExplicit} reduces to
\begin{equation}
    \widetilde P(x,s|x_0)
    =\frac{s+r}{s}\widetilde G_\mu(x,s+r|x_0),
\end{equation}
which is exactly the normalized resetting propagator of Sec.~\ref{sec:reset-no-kill-drift}.

If $r=0$, then Eq.~\eqref{eq:SDriftLaplaceFinal} reduces to $\widetilde S(s)=\widetilde Q_{k,\mu}(s)$, and Eq.~\eqref{eq:PDriftLaplaceCompact} reduces to $\widetilde P(x,s|x_0)=\widetilde q_{k,\mu}(x,s|x_0)$, as it must.

If $\mu=0$, then $\Gamma_r(s)=2\sqrt{D(s+r)}$, and
\begin{equation}
    B_\mu(s)\Big|_{\mu=0}
    =\frac{k\exp[-|x_0|\sqrt{(s+r)/D}]}
    {2\sqrt{D(s+r)}+k},
\end{equation}
which gives the driftless result.

If $k\to\infty$, the delta killing point becomes a perfectly absorbing target at the origin. In that limit
\begin{equation}
    B_\mu(s)\longrightarrow
    \exp\left[-\frac{\mu x_0+\Gamma_r(s)|x_0|}{2D}\right],
\end{equation}
and
\begin{equation}
    \widetilde S_\infty(s)
    =\frac{1-\exp\left[-\frac{\mu x_0+\Gamma_r(s)|x_0|}{2D}\right]}
    {s+r\exp\left[-\frac{\mu x_0+\Gamma_r(s)|x_0|}{2D}\right]}.
    \label{eq:SAbsorbingDriftReset}
\end{equation}
For $\mu=0$, this becomes the standard survival transform for diffusion with stochastic resetting to $x_0$ and a perfectly absorbing target at the origin \cite{EvansMajumdarSchehr2020}.

\subsection{Long-time behavior}
\label{sec:long-time-drift}

We now analyze the large-time behavior. The no-reset and resetting cases are qualitatively different. Without resetting, a nonzero drift makes the motion transient with respect to the origin and partial delta killing leaves a nonzero probability of ultimate survival. With resetting, the particle is repeatedly brought back to $x_0$, and for every $r>0$ and $k>0$ the survival probability decays exponentially to zero.

\subsubsection{No resetting: \texorpdfstring{$r=0$}{r=0}}

The long-time limit of the no-reset survival probability is obtained from the final-value theorem applied to Eq.~\eqref{eq:QDriftLaplace}:
\begin{equation}
    Q_{k,\mu}(\infty)=\lim_{s\to0}s\widetilde Q_{k,\mu}(s).
\end{equation}
For $\mu\ne0$, this gives the reset-free survival plateau
\begin{equation}
    Q_{k,\mu}(\infty)
    =1-\frac{k}{|\mu|+k}
    \exp\left[-\frac{\mu x_0+|\mu||x_0|}{2D}\right].
    \label{eq:QinftyNoResetDrift}
\end{equation}
Thus, unlike driftless Brownian motion, which is recurrent, a drifted Brownian motion with $\mu\neq0$ is not killed with probability one by a delta trap of finite strength $k$.

For example, if $x_0>0$ and $\mu>0$, the drift points away from the delta killing point. Then
\begin{equation}
    Q_{k,\mu}(\infty)
    =1-\frac{k}{\mu+k}\exp\left[-\frac{\mu x_0}{D}\right].
\end{equation}
The factor $\exp[-\mu x_0/D]$ is the usual probability that a drifted Brownian motion starting at $x_0>0$ ever reaches the origin~\cite{Redner,BorodinSalminen}. The additional factor $k/(\mu+k)$ reflects the fact that the killing at the origin is partial, not perfectly absorbing.

If $x_0>0$ and $\mu<0$, the drift points toward the delta killing point. Equation~\eqref{eq:QinftyNoResetDrift} becomes
\begin{equation}
    Q_{k,\mu}(\infty)=1-\frac{k}{|\mu|+k}=\frac{|\mu|}{|\mu|+k}.
\end{equation}
This is nonzero for finite $k$. The reason is that delta killing depends on the accumulated local time at the origin. With nonzero drift, the total local time accumulated at the origin is almost surely finite.

The absence of any dependence on $x_0$ and $D$ in this particular case has a direct interpretation. For $x_0>0$ and $\mu<0$, the origin is reached with probability one. By the strong Markov property at the first hitting time of the origin, the subsequent survival probability is therefore the same as for a particle starting at the trap, so the dependence on $x_0$ disappears. Moreover, at zero Laplace frequency the free return resolvent at the trap is
\begin{equation*}
    \widetilde G_\mu(0,0|0)=\frac{1}{|\mu|},
\end{equation*}
which is independent of $D$. The rank-one formula then gives $\widetilde q_{k,\mu}(0,0|0)=1/(|\mu|+k)$, and integration of the killing flux yields
\begin{equation*}
    Q_{k,\mu}(\infty|0)
    =1-k\widetilde q_{k,\mu}(0,0|0)
    =\frac{|\mu|}{|\mu|+k}.
\end{equation*}
Thus $x_0$ drops out because the trap is reached almost surely before the relevant local dynamics begins, while $D$ drops out because the zero-frequency return weight at the origin is set solely by the drift speed $|\mu|$.

In the limit $k\to\infty$, the delta killing point becomes perfectly absorbing and the survival probability tends to zero when the drift points toward the origin.

The special driftless case $\mu=0$ is singular because Brownian motion is recurrent. Then Eq.~\eqref{eq:QDriftLaplace} gives
\begin{equation}
    Q_{k,\mu}(t)\sim \frac{|x_0|+2D/k}{\sqrt{\pi Dt}},
    \qquad t\to\infty.
    \label{eq:QDriftlessNoResetAsymptotic}
\end{equation}
Thus the survival decays algebraically as $t^{-1/2}$.

For $\mu\ne0$, the approach to the limiting value \eqref{eq:QinftyNoResetDrift} is controlled by the branch point of $\Gamma(s)=\sqrt{\mu^2+4Ds}$ at
\begin{equation}
    s=-\frac{\mu^2}{4D}.
\end{equation}
Consequently,
\begin{equation}
    Q_{k,\mu}(t)-Q_{k,\mu}(\infty)
    =O\left(\mathrm{e}^{-\mu^2t/(4D)}t^{-3/2}\right),
    \qquad t\to\infty,
    \qquad \mu\ne0.
    \label{eq:NoResetDriftCorrection}
\end{equation}
The exponential scale $4D/\mu^2$ is the natural time scale associated with biased diffusion against its drift.

\subsubsection{Resetting: \texorpdfstring{$r>0$}{r>0}}

For $r>0$ and $k>0$, the long-time behavior is governed by the singularity of $\widetilde S(s)$ in Eq.~\eqref{eq:SDriftLaplaceFinal} that lies closest to the origin on the negative real axis. The square root $\Gamma_r(s)=\sqrt{\mu^2+4D(s+r)}$ has a branch point at
\begin{equation*}
    s_{\mathrm b}=-r-\frac{\mu^2}{4D}.
\end{equation*}
Since the physical pole found below satisfies $0<\theta<r$, one has $-\theta>-r\geq s_{\mathrm b}$; the pole therefore lies strictly to the right of the branch point and controls the leading long-time behavior. This singularity is a pole $s=-\theta$, where $\theta>0$ is the asymptotic decay rate. The pole condition is
\begin{equation}
    s+rB_\mu(s)=0
    \quad \text{at} \quad s=-\theta.
\end{equation}
Equivalently,
\begin{equation}
    \theta=rB_\mu(-\theta).
    \label{eq:thetaEquationAbstract}
\end{equation}
Using Eq.~\eqref{eq:BmuDefinition}, define
\begin{equation}
    \Gamma_\theta=\sqrt{\mu^2+4D(r-\theta)}.
    \label{eq:GammaThetaDefinition}
\end{equation}
The central long-time result for the drifted problem is the implicit equation
\begin{equation}
    \theta
    =r\frac{k}{\Gamma_\theta+k}
    \exp\left[-\frac{\mu x_0+\Gamma_\theta |x_0|}{2D}\right].
    \label{eq:thetaEquationExplicit}
\end{equation}

The physical real solution is in fact unique in $0<\theta<r$. To see this, set $u=r-\theta$. From Eq.~\eqref{eq:SDriftLaplaceGeneral}, the pole condition is equivalently
\begin{equation*}
    r\,\widetilde Q_{k,\mu}(u)=1,
    \qquad 0<u<r.
\end{equation*}
Because $Q_{k,\mu}(t)\geq0$, its Laplace transform is strictly decreasing for $u>0$,
\begin{equation*}
    \frac{\mathrm d}{\mathrm du}\widetilde Q_{k,\mu}(u)
    =-\int_0^\infty t\,\mathrm e^{-ut}Q_{k,\mu}(t)\,\mathrm dt<0.
\end{equation*}
Furthermore, $\widetilde Q_{k,\mu}(u)\to\infty$ as $u\downarrow0$ (for $\mu\neq0$ because $Q_{k,\mu}(t)$ approaches the positive plateau above, and for $\mu=0$ because of the $t^{-1/2}$ tail), whereas $r\widetilde Q_{k,\mu}(r)=1-B_\mu(0)<1$. Hence there is exactly one $u\in(0,r)$, and therefore exactly one physical pole $\theta\in(0,r)$. This monotonicity also provides a robust one-dimensional numerical selection criterion for the physical root of Eq.~\eqref{eq:thetaEquationExplicit}.

The finite killing strength enters Eq.~\eqref{eq:thetaEquationExplicit} through the factor
\begin{equation}
    \frac{k}{\Gamma_\theta+k}.
\end{equation}
This factor is absent for a perfectly absorbing target. Hence finite delta killing always slows the asymptotic decay compared with the perfectly absorbing limit, all other parameters being fixed — an expected result, since a target of finite reactivity can only remove probability mass more slowly than one that absorbs on contact.

For weak killing, $k\ll \Gamma(r)$, the pole is close to the origin and one may set $\theta=0$ on the right-hand side to leading order. This gives
\begin{equation}
    \theta
    \simeq r\frac{k}{\Gamma(r)}
    \exp\left[-\frac{\mu x_0+\Gamma(r)|x_0|}{2D}\right],
    \qquad k\ll \Gamma(r).
    \label{eq:WeakKillingThetaMu}
\end{equation}

Using the stationary density of the drifted resetting process, Eq.~\eqref{eq:stationaryResetDrift}, evaluated at the origin,
\begin{equation*}
    p_r^{\mathrm{st}}(0|x_0)
    =\frac{r}{\Gamma(r)}
    \exp\left[-\frac{\mu x_0+\Gamma(r)|x_0|}{2D}\right],
\end{equation*}
Eq.~\eqref{eq:WeakKillingThetaMu} can be written in the same form as in the driftless case,
\begin{equation*}
    \theta\simeq k\,p_r^{\mathrm{st}}(0|x_0).
\end{equation*}
Thus, for weak reactivity, the asymptotic decay rate is again the killing strength times the stationary probability density of the corresponding resetting process at the trap.

For strong killing, $k\to\infty$, Eq.~\eqref{eq:thetaEquationExplicit} becomes
\begin{equation}
    \theta
    =r\exp\left[-\frac{\mu x_0+\Gamma_\theta |x_0|}{2D}\right].
    \label{eq:StrongKillingTheta}
\end{equation}
For $\mu=0$ this reduces to
\begin{equation}
    \theta=r\exp\left[-|x_0|\sqrt{\frac{r-\theta}{D}}\right],
\end{equation}
which is the pole equation for a perfectly absorbing target with resetting~\cite{EvansMajumdarSchehr2020}.

\subsubsection{Residue and asymptotic amplitude}

The pole at $s=-\theta$ is simple. Write
\begin{equation}
    M(s)=s+rB_\mu(s),
    \qquad
    N(s)=1-B_\mu(s).
\end{equation}
Thus
\begin{equation}
    \widetilde S(s)=\frac{N(s)}{M(s)}.
\end{equation}
At the pole,
\begin{equation}
    B_\mu(-\theta)=\frac{\theta}{r},
    \qquad
    N(-\theta)=1-\frac{\theta}{r}.
\end{equation}
The derivative of $B_\mu(s)$ is obtained by differentiating Eq.~\eqref{eq:BmuDefinition}. Since
\begin{equation}
    \frac{\mathrm{d}\Gamma_r}{\mathrm{d}s}
    =\frac{2D}{\Gamma_r},
\end{equation}
one finds
\begin{equation}
    B_\mu'(s)
    =-B_\mu(s)\left[
    \frac{|x_0|}{\Gamma_r(s)}
    +\frac{2D}{\Gamma_r(s)(\Gamma_r(s)+k)}
    \right].
    \label{eq:BDerivative}
\end{equation}
Therefore
\begin{equation}
    M'(-\theta)
    =1+rB_\mu'(-\theta)
    =1-\theta\left[
    \frac{|x_0|}{\Gamma_\theta}
    +\frac{2D}{\Gamma_\theta(\Gamma_\theta+k)}
    \right].
\end{equation}
The residue gives the large-time asymptotics
\begin{equation}
    S(t)\sim A\mathrm{e}^{-\theta t},
    \qquad t\to\infty,
    \label{eq:SurvivalAsymptoticDrift}
\end{equation}
with amplitude
\begin{equation}
    A=\frac{1-\theta/r}
    {1-\theta\left[
    |x_0|/\Gamma_\theta+2D/(\Gamma_\theta(\Gamma_\theta+k))
    \right]}.
    \label{eq:AmplitudeDrift}
\end{equation}
The denominator is the slope of the pole equation in Laplace space. When $k\to\infty$, the term involving $\Gamma_\theta+k$ drops out and the expression reduces to the perfectly absorbing target result. When $\mu=0$, $\Gamma_\theta=2\sqrt{D(r-\theta)}$, and Eq.~\eqref{eq:AmplitudeDrift} becomes the driftless amplitude.

\subsubsection{Asymptotic density and quasi-stationary shape}

The same pole controls the non-normalized propagator. From Eq.~\eqref{eq:PDriftLaplaceCompact}, near $s=-\theta$,
\begin{equation}
    \widetilde P(x,s|x_0)
    \simeq
    \frac{(r-\theta)\widetilde q_{k,\mu}(x,r-\theta|x_0)}
    {M'(-\theta)}\frac{1}{s+\theta}.
\end{equation}
Thus
\begin{equation}
    P(x,t|x_0)\sim A p_{\mathrm{qs}}(x)\mathrm{e}^{-\theta t},
    \qquad t\to\infty,
    \label{eq:PAsymptoticDrift}
\end{equation}
where the conditional quasi-stationary density is
\begin{equation}
    p_{\mathrm{qs}}(x)=r\widetilde q_{k,\mu}(x,r-\theta|x_0).
    \label{eq:QSDCompact}
\end{equation}
It is normalized because the pole condition is equivalent to
\begin{equation}
    r\widetilde Q_{k,\mu}(r-\theta)=1.
\end{equation}
Using Eq.~\eqref{eq:qDriftLaplace}, the quasi-stationary density is explicitly
\begin{equation}
    \begin{aligned}
    p_{\mathrm{qs}}(x)
    =&\frac{r}{\Gamma_\theta}
    \bigg\{
    \exp\left[\frac{\mu(x-x_0)-\Gamma_\theta|x-x_0|}{2D}\right]
    \\
    &\hspace{1.7cm}
    -\frac{k}{\Gamma_\theta+k}
    \exp\left[\frac{\mu(x-x_0)-\Gamma_\theta(|x|+|x_0|)}{2D}\right]
    \bigg\}.
    \end{aligned}
    \label{eq:QSDExplicit}
\end{equation}
This density describes the spatial distribution of the particle conditioned on survival at long times. It is asymmetric when $\mu\ne0$. 

\begin{figure}[t]
\centering
\includegraphics[width=0.88\columnwidth]{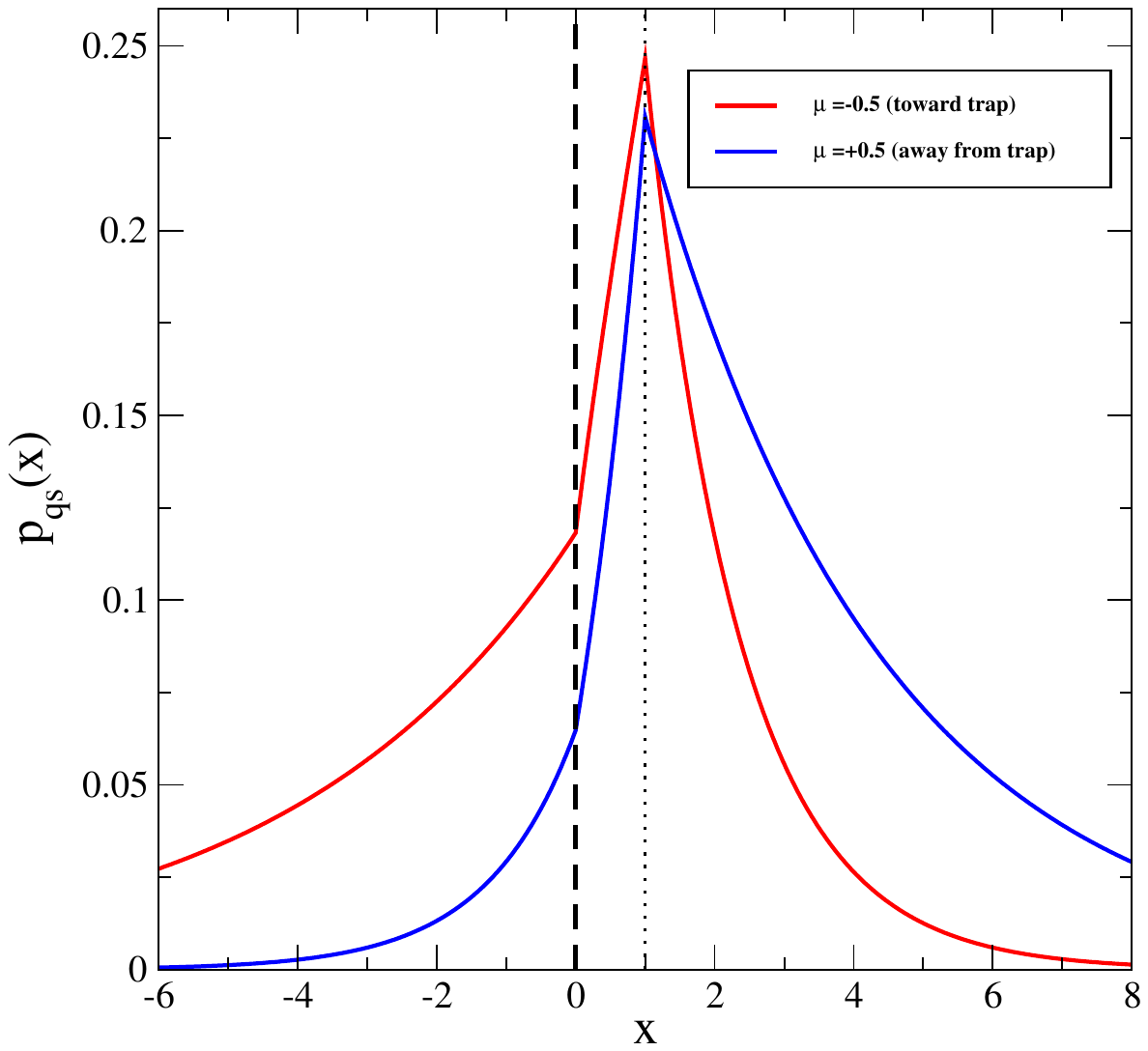}
\caption{Exact quasi-stationary density $p_{\mathrm{qs}}(x)$ from Eq.~\eqref{eq:QSDExplicit} for $D=1$, $x_0=1$, $k=1$, and $r=0.3$. The red curve corresponds to $\mu=-0.5$ (drift toward the trap), while the blue curve corresponds to $\mu=+0.5$ (drift away from the trap). The dashed vertical line marks the trap at $x=0$, and the dotted vertical line marks the reset position $x_0=1$. The two profiles are both normalized and clearly display the drift-induced asymmetry of the conditioned long-time state.}
\label{fig:qsd-drift}
\end{figure}

Figure~\ref{fig:qsd-drift} illustrates the drift-induced asymmetry of the quasi-stationary state by comparing profiles for equal drift magnitudes directed toward and away from the trap. Although the two cases differ only by the sign of the drift, their conditioned long-time spatial profiles are markedly different.

The jump condition at the origin remains
\begin{equation}
    D\left[p_{\mathrm{qs}}'(0^+)-p_{\mathrm{qs}}'(0^-)\right]
    =kp_{\mathrm{qs}}(0).
\end{equation}
Moreover, Eq.~\eqref{eq:qOriginDriftLaplace} evaluated at $s=r-\theta$ gives
\begin{equation}
    kp_{\mathrm{qs}}(0)=\theta.
\end{equation}
Thus the exponential decay rate is exactly the killing flux of the quasi-stationary state.

\subsubsection{Physical interpretation}
 
The role of resetting is to turn each excursion after a reset into a statistically renewed cycle. During one cycle, the particle starts from $x_0$, drifts and diffuses, and may be killed through the local time accumulated at the origin. The factor $B_\mu(s)$ is the Laplace-space weight of absorption during one reset-free cycle. The denominator
\begin{equation}
    s+rB_\mu(s)
\end{equation}
resums the infinite sequence of such cycles. Its zero determines the long-time decay rate.

The drift enters $B_\mu(s)$, defined in Eq.~\eqref{eq:BmuDefinition},
through two physically distinct effects. First, the exponential factor
\begin{equation}
    \exp\left[-\frac{\mu x_0+\Gamma_r(s)|x_0|}{2D}\right]
\end{equation}
is the Laplace transform of the first-passage-time density from $x_0$ to the
origin. Being a function of $s$, it is not itself a probability for
general $s$. At $s=0$, however, the exponential discount associated with
the resetting rate gives it the direct interpretation of the probability
that the particle reaches the origin before the next reset.
This factor alone carries the sign of $\mu x_0$: when $\mu x_0>0$,
the drift pushes the particle away from the origin and this factor suppresses
the reaching probability; when $\mu x_0<0$, the drift points toward the
origin and the reaching probability is enhanced.

Second, given that the origin has been reached, the prefactor
$k/(\Gamma_r(s)+k)$ is, in the same sense, the Laplace transform,
evaluated at $s+r$, of the absorption-time density for a particle starting
at the origin. At $s=0$, it becomes the conditional probability that the
particle is killed before the next reset, given that the origin has been
reached. This second factor depends on $\mu$ only through
$\Gamma_r(s)=\sqrt{\mu^2+4D(s+r)}$, i.e., symmetrically in the sign of the
drift: once the particle sits at the trap, how much local time it accumulates
before escaping depends on the strength of the drift, not on which side of
the origin it came from.

This cycle interpretation also explains why the same pole controls both the
survival probability and the conditioned long-time spatial profile.

\section{Discussion and conclusion}

The addition of a constant drift preserves the renewal structure of stochastic resetting while modifying the reset-free Green functions through the quantity $\Gamma(s)=\sqrt{\mu^2+4Ds}$ --- which has the same dimensions as the drift $\mu$, namely a speed, and reduces to $2\sqrt{Ds}$ at $\mu=0$ --- and the orientation-dependent factor involving $\mu x_0$. The exact combined solution is summarized by Eqs.~\eqref{eq:SDriftLaplaceFinal} and \eqref{eq:PDriftLaplaceExplicit}; its long-time behavior follows from the dominant pole in Eq.~\eqref{eq:thetaEquationExplicit} and the quasi-stationary density in Eq.~\eqref{eq:QSDExplicit}. The driftless and perfectly absorbing limits follow smoothly from these expressions.
 
The combined results answer the physical question posed in the Introduction. The cycle factorization discussed above makes the mechanism explicit: the sign of $\mu x_0$ controls the target-reaching part of each excursion, whereas the finite-reactivity part depends on the magnitude of the drift through local-time accumulation. Resetting removes the survival plateau created by drift-induced transience because the particle is repeatedly returned to a position from which another reactive encounter is possible. The imperfect nature of the target survives in the value of the pole $\theta$ and in the quasi-stationary profile, but not in the qualitative fact of eventual absorption when $r>0$ and $k>0$.
 
The model therefore displays three distinct long-time regimes. For $r=0$ and $\mu=0$, recurrence of one-dimensional Brownian motion produces the algebraic law $S(t)\sim t^{-1/2}$. For $r=0$ and $\mu\neq0$, transience produces a survival plateau, namely the strictly positive limit $S(\infty)=Q_{k,\mu}(\infty)$ given by Eq.~\eqref{eq:QinftyNoResetDrift}. For $r>0$, renewal of statistically independent excursions suppresses that plateau and yields the exponential law $S(t)\sim A\mathrm{e}^{-\theta t}$; the same dominant pole determines the quasi-stationary spatial profile. The orientation of the drift relative to the trap — equivalently, the sign of $\mu x_0$ — enters quantitatively through the probability of successful reactive contact during a reset-free excursion. Thus recurrence, transience, and renewal generate three different survival mechanisms within the same localized-reaction model.

%%%%%%%%%%%%%%%%%%%%%%%%%%%%%%%%

\appendix
\section{Optimal resetting rate for a delta killing trap}
\label{app:optimal-resetting-rate}

This technical appendix revisits the optimization of the mean absorption time with respect to the resetting rate, first analyzed for a partially absorbing target by Whitehouse, Evans, and Majumdar in Sec.~IV A of Ref.~\cite{Whitehouse2013}. We regard the mean absorption time in Eq.~\eqref{eq:MeanLifetime} as a function of $r>0$, while holding $D>0$, $k>0$, and the reset-to-trap distance $|x_0|>0$ fixed:
\begin{equation}
    T(r)
    =
    \frac{1}{r}
    \left[
        \left(1+\frac{2\sqrt{Dr}}{k}\right)
        \exp\left(|x_0|\sqrt{\frac{r}{D}}\right)-1
    \right].
    \label{eq:appendix-mean-time}
\end{equation}
Whitehouse \emph{et al.} derived the corresponding stationarity condition and analyzed the strong- and weak-absorption limits. Building on their calculation, we recast the optimization in terms of a dimensionless reactivity parameter that is independent of $r$. This choice makes the minimization at fixed physical parameters explicit and allows us to give a direct proof that $T(r)$ possesses a unique global minimum for every finite $k>0$. It also provides a convenient form for studying how the optimal rate varies with the trap reactivity.

\subsection{Dimensionless formulation}

Introduce the dimensionless variables
\begin{equation}
    z=|x_0|\sqrt{\frac{r}{D}},
    \qquad
    \eta=\frac{2D}{|x_0|k}.
    \label{eq:appendix-dimensionless-variables}
\end{equation}
The variable $z$ is the reset-to-trap distance measured in units of the diffusive length $\sqrt{D/r}$, whereas $\eta$ measures the imperfection of the trap. The perfectly absorbing limit $k\to\infty$ corresponds to $\eta\to0$, while $\eta\gg1$ describes weak reactivity. Importantly, $\eta$ remains fixed when $r$ is varied at fixed $D$, $|x_0|$, and $k$. In the notation of Ref.~\cite{Whitehouse2013}, the present variables satisfy $z=\theta$ and $\eta z=2\phi_0$.

Since
\begin{equation}
    r=\frac{D}{|x_0|^2}z^2,
\end{equation}
Eq.~\eqref{eq:appendix-mean-time} becomes
\begin{equation}
    T(r)=\frac{|x_0|^2}{D}\,\tau_\eta(z),
    \qquad
    \tau_\eta(z)
    =
    \frac{(1+\eta z)\mathrm{e}^{z}-1}{z^2}.
    \label{eq:appendix-dimensionless-time}
\end{equation}
Thus, for fixed $\eta$, minimizing $T(r)$ over $r>0$ is equivalent to minimizing $\tau_\eta(z)$ over $z>0$.

The endpoint behavior guarantees the existence of at least one global minimizer. As $z\to0$, equivalently $r\to0$,
\begin{equation}
    \tau_\eta(z)
    =
    \frac{1+\eta}{z}
    +\left(\frac{1}{2}+\eta\right)
    +O(z),
\end{equation}
and hence
\begin{equation}
    T(r)
    =
    \left(
        \frac{|x_0|}{\sqrt{D}}+\frac{2\sqrt{D}}{k}
    \right)\frac{1}{\sqrt{r}}
    +
    \frac{|x_0|^2}{2D}
    +\frac{2|x_0|}{k}
    +O(\sqrt{r}).
    \label{eq:appendix-small-r}
\end{equation}
Therefore $T(r)\to+\infty$ as $r\to0$. At the opposite endpoint,
\begin{equation}
    T(r)
    \sim
    \frac{1}{r}
    \left(1+\frac{2\sqrt{Dr}}{k}\right)
    \exp\left(|x_0|\sqrt{\frac{r}{D}}\right)
    \qquad (r\to\infty),
\end{equation}
so $T(r)\to+\infty$ also in the limit of very frequent resetting. Continuity then ensures that $T(r)$ attains at least one minimum at a finite positive value of $r$.

\subsection{Stationary condition and uniqueness}

Differentiating Eq.~\eqref{eq:appendix-dimensionless-time} gives
\begin{equation}
    \tau_\eta'(z)
    =
    \frac{H_\eta(z)}{z^3},
\end{equation}
where
\begin{equation}
    H_\eta(z)
    =
    \mathrm{e}^{z}
    \left[
        \eta z^2+(1-\eta)z-2
    \right]+2.
    \label{eq:appendix-H}
\end{equation}
The optimal value $z_*$ is consequently determined by the transcendental equation
\begin{equation}
    \mathrm{e}^{z_*}
    \left[
        \eta z_*^2+(1-\eta)z_*-2
    \right]+2=0.
    \label{eq:appendix-optimal-equation}
\end{equation}
The corresponding optimal resetting rate is
\begin{equation}
    r_*=\frac{D}{|x_0|^2}z_*^2.
    \label{eq:appendix-optimal-rate}
\end{equation}

The stationary point is unique. Indeed,
\begin{equation}
    H_\eta'(z)
    =
    \mathrm{e}^{z}
    \left[
        \eta z^2+(1+\eta)z-(1+\eta)
    \right].
    \label{eq:appendix-Hprime}
\end{equation}
The polynomial in square brackets is strictly increasing for $z>0$
and has a single positive zero. Thus $H_\eta$ first decreases and
then increases. Moreover,
\begin{equation}
    H_\eta(1)=2-\mathrm{e}<0,
    \qquad
    H_\eta(2)=2+2\eta\mathrm{e}^{2}>0.
\end{equation}
It follows that Eq.~\eqref{eq:appendix-optimal-equation} has exactly
one solution in the interval
\begin{equation}
    1<z_*<2.
\end{equation}
Since the mean time diverges at both endpoints, this stationary point
is the unique global minimum.

\subsection{Perfectly absorbing limit}

In the limit $k\to\infty$, or $\eta\to0$,
Eq.~\eqref{eq:appendix-optimal-equation} reduces to
\begin{equation}
    \mathrm{e}^{z_\infty}(z_\infty-2)+2=0,
\end{equation}
or equivalently
\begin{equation}
    z_\infty=2\left(1-\mathrm{e}^{-z_\infty}\right).
    \label{eq:appendix-Evans-Majumdar-equation}
\end{equation}
Besides the trivial solution at the boundary $z=0$, the relevant
positive solution is
\begin{equation}
    z_\infty\simeq1.59362426.
\end{equation}
This is precisely the optimal dimensionless resetting parameter found
by Evans and Majumdar for a perfectly absorbing target in their
foundation article on diffusion with stochastic
resetting~\cite{EvansMajumdar2011}. Therefore,
\begin{equation}
    r_*^{(\infty)}
    =
    \frac{D}{|x_0|^2}z_\infty^2
    \simeq
    2.539638\,\frac{D}{|x_0|^2},
    \label{eq:appendix-perfect-optimal-rate}
\end{equation}
and the corresponding minimum mean absorption time is
\begin{equation}
    T_{\min}^{(\infty)}
    =
    \frac{|x_0|^2}{D}
    \frac{\mathrm{e}^{z_\infty}-1}{z_\infty^2}
    \simeq
    1.544139\,\frac{|x_0|^2}{D}.
    \label{eq:appendix-perfect-minimum-time}
\end{equation}

\subsection{Effect of finite reactivity}

Implicit differentiation of
Eq.~\eqref{eq:appendix-optimal-equation} yields
\begin{equation}
    \frac{\mathrm{d}z_*}{\mathrm{d}\eta}
    =
    -
    \frac{z_*(z_*-1)}
    {\eta z_*^2+(1+\eta)z_*-(1+\eta)}.
    \label{eq:appendix-monotonicity}
\end{equation}
The denominator is positive at the unique crossing of $H_\eta$, and
$z_*>1$. Consequently,
\begin{equation}
    \frac{\mathrm{d}z_*}{\mathrm{d}\eta}<0.
\end{equation}
Since $\eta$ increases when $k$ decreases, a less reactive trap shifts
the optimum toward a smaller resetting rate. In particular,
\begin{equation}
    1<z_*(\eta)\leq z_\infty,
\end{equation}
Combining these inequalities yields the following bounds on the optimal resetting rate:
\begin{equation}
    \frac{D}{|x_0|^2}
    <
    r_*
    \leq
    2.539638\,\frac{D}{|x_0|^2}.
    \label{eq:appendix-rate-bounds}
\end{equation}
The upper bound is attained in the perfectly absorbing limit.

At the optimum, the condition
$z_*N'(z_*)=2N(z_*)$, where
\begin{equation}
    N(z)=(1+\eta z)\mathrm{e}^{z}-1,
\end{equation}
gives a compact expression for the minimum mean absorption time:
\begin{equation}
    T_{\min}
    =
    \frac{|x_0|^2}{D}
    \frac{
        \mathrm{e}^{z_*}
        \left[1+\eta(1+z_*)\right]
    }{2z_*}.
    \label{eq:appendix-minimum-time}
\end{equation}
Differentiating the optimized dimensionless mean absorption time, one obtains
\begin{align}
    \frac{\mathrm{d}}{\mathrm{d}\eta}
    \tau_\eta\bigl(z_*(\eta)\bigr)
    &={}
    \left.\frac{\partial \tau_\eta}{\partial \eta}\right|_{z=z_*}
    +
    \left.\frac{\partial \tau_\eta}{\partial z}\right|_{z=z_*}
    \frac{\mathrm{d}z_*}{\mathrm{d}\eta}
    \notag\\
    &={}
    \left.\frac{\partial \tau_\eta}{\partial \eta}\right|_{z=z_*}
    =\frac{\mathrm{e}^{z_*}}{z_*}>0.
    \label{eq:appendix-envelope}
\end{align}
In the second line we used the stationarity condition
$\partial_z\tau_\eta(z_*)=0$. This elementary cancellation is the
smooth finite-dimensional form of the envelope theorem
\cite{MilgromSegalEnvelope}. It follows that the minimum mean
absorption time increases monotonically as the trap becomes less
reactive.

Finally, in the weak-reactivity limit $k\to0^+$, or
$\eta\to\infty$, Eq.~\eqref{eq:appendix-optimal-equation} gives
\begin{equation}
    z_*
    =
    1+\frac{1-2/\mathrm{e}}{\eta}
    +O(\eta^{-2}).
    \label{eq:appendix-weak-reactivity-z}
\end{equation}
It follows that
\begin{equation}
    r_*\longrightarrow\frac{D}{|x_0|^2},
    \qquad k\to0^+,
\end{equation}
whereas the optimized mean absorption time diverges as
\begin{equation}
    T_{\min}
    =
    \frac{2\mathrm{e}|x_0|}{k}
    +(\mathrm{e}-1)\frac{|x_0|^2}{D}
    +O(k).
    \label{eq:appendix-weak-reactivity-time}
\end{equation}
The finite limit of $r_*$ should therefore not be confused with a
finite absorption time: as the trap becomes nonreactive, the optimal
rate remains of order $D/|x_0|^2$, but the minimum mean absorption time
diverges as $k^{-1}$.

\subsection{Effect of a constant drift}

The analysis above assumed $\mu=0$. For a weak drift, the leading effect on
the mean absorption time follows from expanding Eq.~\eqref{eq:MeanLifetimeDrift}
to first order in $\mu$ at fixed $r$:
\begin{equation}
    T_\mu(r)
    =
    T(r)
    +
    \frac{\mu x_0}{2D}
    \left[
        T(r)+\frac{1}{r}
    \right]
    +O(\mu^2),
    \label{eq:appendix-weak-drift-T}
\end{equation}
where $T(r)$ is the driftless mean absorption time of
Eq.~\eqref{eq:appendix-mean-time}. Let $r_0^*$ denote the unique optimal
resetting rate in the driftless problem established above. By the same
envelope-theorem argument used in Eq.~\eqref{eq:appendix-envelope}, the
leading correction to the minimum absorption time is
\begin{equation}
    \left.
    \frac{\mathrm{d}T_{\min}}{\mathrm{d}\mu}
    \right|_{\mu=0}
    =
    \frac{x_0}{2D}
    \left[
        T(r_0^*)+\frac{1}{r_0^*}
    \right].
\end{equation}
For $x_0>0$, this derivative is positive: to first order in the drift, a
drift away from the trap increases the minimum mean absorption time, whereas
a drift toward the trap decreases it.

Applying the implicit function theorem to the first-order condition
$\partial_r T_\mu(r)=0$ at $r=r_0^*$, and using $T'(r_0^*)=0$, gives the
corresponding shift of the optimal resetting rate:
\begin{equation}
    \left.
    \frac{\mathrm{d}r_*(\mu)}{\mathrm{d}\mu}
    \right|_{\mu=0}
    =
    \frac{x_0}
    {2D(r_0^*)^2 T''(r_0^*)},
    \qquad (x_0>0).
    \label{eq:appendix-drift-rate-shift}
\end{equation}
Since $T''(r_0^*)>0$ at the unique global minimum established above, this
derivative is positive. Thus, near $\mu=0$, drift toward the trap
($\mu x_0<0$) lowers the optimal resetting rate below its driftless value
$r_0^*$, whereas drift away from the trap ($\mu x_0>0$) raises it above
$r_0^*$. Less frequent resetting is needed when the drift itself assists the
particle in reaching the target, while more frequent resetting is needed to
counteract a drift that carries the particle away.

The exact drifted mean absorption time in Eq.~\eqref{eq:MeanLifetimeDrift}
also permits a compact extension of the optimization variables introduced
above, valid for arbitrary drift strength and not only near $\mu=0$. Define
\begin{equation*}
    m=\frac{|\mu||x_0|}{2D},
    \qquad
    \sigma=\operatorname{sgn}(\mu x_0),
    \qquad
    w(z)=\sqrt{m^2+z^2},
\end{equation*}
while retaining
\begin{equation*}
    z=|x_0|\sqrt{\frac{r}{D}},
    \qquad
    \eta=\frac{2D}{|x_0|k}.
\end{equation*}
The resetting rate and the dimensionless variable $z$ are therefore related
one-to-one by
\begin{equation*}
    r=\frac{D}{|x_0|^2}z^2.
\end{equation*}
Using these definitions, the drifted mean absorption time can be written as
\begin{equation}
    T_\mu\!\left(\frac{Dz^2}{|x_0|^2}\right)
    =
    \frac{|x_0|^2}{D}\,
    \tau_{\eta,m,\sigma}(z),
    \qquad
    \tau_{\eta,m,\sigma}(z)
    =
    \frac{
        \left[1+\eta w(z)\right]
        \exp\left[\sigma m+w(z)\right]-1
    }{z^2}.
    \label{eq:appendix-drift-dimensionless-time}
\end{equation}
Indeed,
\begin{equation*}
    \frac{\mu x_0}{2D}
    =
    \sigma m,
    \qquad
    \frac{|x_0|\Gamma(r)}{2D}
    =
    \sqrt{m^2+z^2}
    =
    w(z),
    \qquad
    \frac{\Gamma(r)}{k}
    =
    \eta w(z),
\end{equation*}
which gives Eq.~\eqref{eq:appendix-drift-dimensionless-time} directly from
Eq.~\eqref{eq:MeanLifetimeDrift}.

Since the map
\begin{equation*}
    r\mapsto z=|x_0|\sqrt{\frac{r}{D}}
\end{equation*}
is one-to-one and strictly increasing from $r>0$ to $z>0$, minimizing
$T_\mu(r)$ with respect to the resetting rate is exactly equivalent to
minimizing $\tau_{\eta,m,\sigma}(z)$ over $z>0$.

For fixed drift magnitude $m$, reversing the drift changes only the factor
$\exp(\sigma m)$, but this is sufficient to shift the minimizer of
$\tau_{\eta,m,\sigma}(z)$. Thus the optimal resetting rate is generally
sensitive to whether the drift points toward or away from the trap, even
when $|\mu|$ is unchanged. The minimization remains one-dimensional and can
be performed directly from
Eq.~\eqref{eq:appendix-drift-dimensionless-time}.
For the representative parameters $D=1$, $k=2$, and $x_0=1$ used in
Fig.~\ref{fig:mean-absorption-drift}, the minima occur at
$r_*\simeq1.50$ for $\mu=-1$, $r_0^*\simeq1.38$ for $\mu=0$, and
$r_*\simeq2.06$ for $\mu=1$.

In fact, for these parameters, $r_*(\mu)$ is non-monotonic on the
drift-toward-trap side: as $\mu$ decreases below zero, the optimal rate first
falls below its driftless value, reaches a minimum
$r_*\simeq1.30$ at $\mu\simeq-0.37$, and then increases again for stronger
negative drift. Physically, a sufficiently strong drift carries the particle
rapidly through the trap region, reducing the local time accumulated during
each encounter and thereby making finite reactivity increasingly important.

Physically, the optimum reflects a competition between two effects.
Resetting suppresses exceptionally long diffusive excursions, but
excessively frequent resetting repeatedly interrupts trajectories
before they can reach the trap and accumulate enough local time to be
absorbed. Finite reactivity strengthens the latter effect and
therefore shifts the optimal resetting rate below the perfect-target
value found by Evans and Majumdar~\cite{EvansMajumdar2011}.

\section*{Data Availability Statement}
There are no publicly available research data or software supporting this manuscript. Requests for further information or data should be sent to the author.

\end{document}